\documentclass[10pt]{article}
\usepackage[utf8]{inputenc}
\usepackage[T1]{fontenc}
\usepackage{amsmath, amssymb, amsfonts}
\usepackage{geometry}
\usepackage{graphicx}
\usepackage{algorithm}
\usepackage{algpseudocode}
\usepackage{float}
\usepackage{subcaption}
\usepackage{physics}
\usepackage{booktabs} 
\usepackage[
    colorlinks=true,
    linkcolor=blue,
    citecolor=blue,
    urlcolor=cyan
]{hyperref}
\usepackage{amsthm}

\DeclareUnicodeCharacter{200B}{}
\newcommand{\uvec}{\mathbf{u}}

\newcommand{\Lvec}{\mathbf{L}}
\newcommand{\Lphi}{\mathbf{L}_\Phi}
\newcommand{\LA}{\mathbf{L}_A}
\newcommand{\PhiL}{\Phi_L}

\newcommand{\AL}{\mathbf{A}_L}
\newcommand{\omvec}{\boldsymbol{\omega}}

\newcommand{\Ttransfer}{T_{\Phi \to A}}
\newcommand{\Ephi}{E_\Phi}
\newcommand{\EA}{E_A}
\newcommand{\EL}{E_L}

\newcommand{\hatAL}{\hat{\mathbf{A}}_L}
\newcommand{\hatEPhi}{\hat{E}_\Phi}
\newcommand{\hatEA}{\hat{E}_A}
\newcommand{\hatLPhi}{\hat{L}_\Phi}
\newcommand{\hatLA}{\hat{L}_A}
\newcommand{\hatTtransfer}{\hat{T}_{\Phi \to A}}

\newcommand{\Zcal}{\mathcal{Z}}
\newcommand{\Rcal}{\mathcal{R}}
\newcommand{\rvec}{\mathbf{r}}
\newcommand{\xvec}{\mathbf{x}}
\newcommand{\kvec}{\mathbf{k}}
\newcommand{\Avec}{\mathbf{A}}
\newcommand{\Bvec}{\mathbf{B}}
\newcommand{\Cvec}{\mathbf{C}}
\newcommand{\Fvec}{\mathbf{F}}

\title{A Statistical Field Theory for Isotropic Turbulence}
\author{Ahmed Farooq\\
University of New Brunswick}
\date{}

\begin{document}

\maketitle

\begin{abstract}
This article establishes a first-principles statistical field theory of fully developed isotropic turbulence. Applying an exact Helmholtz decomposition to the local angular momentum field ($\Lvec = \rvec \times \uvec$) reveals a  segregation into two orthogonally distinct topological phases: a longitudinal condensate of macroscopic coherent structures ($\PhiL$) and a volume-filling, transverse thermal bath ($\AL$). Constructing a Hamiltonian and evaluating the  partition function of these decoupled fields demonstrates that their ergodic exploration of phase space is topologically quantized, mandating a strict $1:2$ equipartition of degrees of freedom. Inverting this topological projection back to the velocity domain isolates the radial velocity field ($\uvec_r$) (which strictly resides in the null space of the $\Lvec$ framework) revealing a recursive partitioning scheme across the cascade into a precise $1/3 : 2/9 : 4/9$ fractional hierarchy. This geometric constraint forces the turbulent steady state into a rigorous canonical equilibrium governed by the equalization of phase chemical potentials ($\mu_\Phi = \mu_A$). The radial component acts as a non-equilibrium mechanical piston, continuously injecting energy into the tangential modes to sustain the canonical equilibrium---a mechanism that mathematically formalizes the classical phenomenology of vortex stretching. Spectral evaluations from direct numerical simulation strongly corroborate this thermodynamic framework, establishing the universality of the partition ratios $1:2$ and $1/3 : 2/9 : 4/9$ as a fundamental signature of three-dimensional isotropic turbulence.
\end{abstract}

\section{Introduction}

Although the statistical approach to turbulence research began with Taylor and Richardson who envisioned an energy cascade as a hierarchical process of eddy breakdown leading to heat dissipation, it was Kolmogorov \cite{kolmogorov1941local} who was able to obtain for the first time a theory and scaling for the energy cascade.  Batchelor \cite{batchelor1953theory} advanced this theory further and also developed a theory of isotropic turbulence. This (statistical) approach has matured into a theory of spectral closure schemes (see Lesieur  \cite{lesieur2008turbulence},  Sagaut \& Cambon \cite{sagaut2018homogeneous}, Verma \cite{verma2019energy} among others). This classical approach, centered on the velocity and vorticity fields provides a powerful, local description of the flow kinematics and dynamics.  \\

However, detailed experimental work (Kline et al. \cite{kline1967structure} for example) revealed the existence of organized, persistent features of the flow---coherent structures---which purely statistical theories cannot account for. The challenge that arose was how to characterize these structures mathematically, see Cantwell \cite{cantwell1981organized}, Robinson \cite{robinson1991review} for example which spurred the development of mathematical extraction schemes, ranging from Lumley's statistical Proper Orthogonal Decomposition (POD) \cite{lumley1967structure} and also to local kinematic identifiers (the $Q$-criterion Hunt et al. \cite{hunt1988eddies}, the $\lambda_2$-criterion Jeong et al. \cite{jeong1995identification}). With these tools, it became possible to isolate specific flow topologies, such as the hairpin vortices characterized by Adrian et al. \cite{adrian2000vortex}. For a comprehensive modern perspective on the dynamic role of these structures, see Jiménez \cite{jimenez2018coherent}.\\

The development of highly accurate spectral methods, (see Orszag and collaborators \cite{gottlieb1971numerical}) for the Direct Numerical Simulation (DNS) of the full Navier-Stokes equations enabled the resolution of full, three-dimensional kinematics of coherent structures. Simulations of wall-bounded flows, such as the channel flow simulation by Kim, Moin, and Moser \cite{kim1987turbulence}, showed the dynamics of near-wall structural dynamics.  Similarly simulations of homogeneous isotropic turbulence, such as those by Vincent et al. \cite{vincent1991spatial}, revealed the spontaneous emergence of intense, tube-like vortex filaments (``worms''). To manage the computational cost of DNS, Large Eddy Simulation (LES) was developed to filter out the small scale dynamics see \cite{smagorinsky1963general, deardorff1970numerical, leonard1975energy, ferziger1977large}, etc.  Challenges still remain at  high Reynolds-number wall turbulence (Smits \cite{smits2011high})  and decoding the effects like compressibility (see Bradshaw \cite{bradshaw1977compressible}, Lele \cite{lele1994compressibility}).\\

Relying on statistical mechanics, it was Onsager \cite{onsager1949statistical} who first proposed a theory for the self-organization of vortices, predicting the existence of negative temperature states. This thermodynamic perspective was later built upon by Miller \cite{miller1990statistical} and Robert \& Sommeria \cite{robert1991statistical}, who applied the entropy maximization principle to predict the final coherent states of 2D turbulence. Parallel to this, Kraichnan \cite{kraichnan1967inertial} established the spectral consequences of these constraints, formulating the dual cascade theory that distinguishes the energy-conserving 3D dynamics from the enstrophy-dominated 2D regime.  Arnold \cite{arnold1966geometrie}  developed the geometric approach to fluid flow via a non-canonical bracket and an energy based Hamiltonian to derive the Euler equations motion. This suggested that the conservation laws are deeply tied to the underlying symmetry groups of the field variables and it is in this spirit that the present work seeks to develop a field-theoretic framework.\\

While the velocity-vorticity framework can be found in standard references (see  Batchelor \cite{batchelor1967introduction}, Aris \cite{aris1989vectors}, Landau and Lifshitz \cite{landau1987fluid}, among many other excellent works) the work presented here introduces a complementary perspective by promoting the specific angular momentum defined as, $\Lvec = \rvec \times \uvec$, to a primary field variable, see Farooq \cite{Farooq1}.  The angular momentum $\Lvec$ framework, which is a non-local, position-weighted quantity, will reveal new features and structure in the underlying equations. \\

This work aims to show that the turbulent field can be  understood through a change of coordinates—from the local velocity vector to the non-local specific angular momentum, $\Lvec $. With this shift in perspective a geometric quantization  and partitioning of the flow energies emerges naturally.\\

The paper is organized as follows. In Section 1-2, we introduce the $\Lvec$ framework and the Helmholtz-Hodge decomposition of $\Lvec$, separating the flow into a coherent, irrotational field ($\Phi_L$) and a background, solenoidal field ($\AL$). In Section 3, we construct the statistical mechanics of the theory. Section 4 maps these potentials back to the physical velocity space, revealing a recursive partitioning of the velocity field.  Section 5 explores the dynamics of this system via the coupled energy budgets. In Section 6 we review key findings of Kolomogorov's theory in the light of the new geometric structuring of turbulnce. In Section 7, we put these predictions to a  test against a large ensemble of Direct Numerical Simulation (DNS) snapshots of 3D isotropic turbulence. Section 8 looks at the stability of the equipartition ratio 1:2 between ($\Phi_L, \AL$) fields in the presence of viscosity using Mean Field Theory.  In Section 9 we define the thermodynamic state of the system and show that the isotropic turbulent relaxes into a two "fluid" equilibrium by a process of thermalization.

\subsection{The Governing Equations}

We begin by characterizing the specific angular momentum field.  To derive the transport equation for the specific angular momentum, $\Lvec$ (also see Farooq \cite{Farooq1}), we start with the incompressible  Navier-Stokes equations:\\

\begin{equation}
\left(\rho \frac{D\uvec}{Dt}\right) = (-\nabla p) + (\mu \nabla^2 \uvec)
\label{eq:moment_of_ns}
\end{equation}

\begin{equation}
\nabla \cdot \uvec = 0
\end{equation}

where $\uvec$ is the velocity field,  $p$ is the pressure field, $\mu$ is the viscosity and $\rho$ is the density.\\

To find a governing equation for the specific angular momentum, $\Lvec$, we take the moment of the Navier-Stokes equation with respect to the origin.\\

\begin{equation}
\rvec \times \left(\rho \frac{D\uvec}{Dt}\right) = \rvec \times (-\nabla p) + \rvec \times (\mu \nabla^2 \uvec)
\label{eq:moment_of_ns}
\end{equation}

The term on the left-hand side can be related to the material derivative of $\Lvec$ using the product rule. The material derivative of the position vector of a fluid parcel is its velocity, $D\rvec/Dt = \uvec$. Therefore:

\begin{align}
\frac{D\Lvec}{Dt} = \frac{D}{Dt}(\rvec \times \uvec) &= \frac{D\rvec}{Dt} \times \uvec + \rvec \times \frac{D\uvec}{Dt} \nonumber = \rvec \times \frac{D\uvec}{Dt}
\end{align}

This provides a direct identity for the left-hand side of Eq. \ref{eq:moment_of_ns}. For the viscous term, we use the vector identity\footnote{To show $\rvec \times (\nabla^2 \uvec) = \nabla^2 (\rvec \times \uvec) - 2 (\nabla \times \uvec)$, let  $\rvec = (x_1, x_2, x_3)$, $\uvec = (u_1, u_2, u_3)$. The left-hand side is: $(\rvec \times (\nabla^2 \uvec))_i = \epsilon_{ijk} x_j (\nabla^2 u_k) = \epsilon_{ijk} x_j \partial_m \partial_m u_k$.  The right-hand side involves $\Lvec$, where $L_i = \epsilon_{ijk} x_j u_k$. Compute the Laplacian: $ \nabla^2 L_i = \partial_m \partial_m (\epsilon_{ijk} x_j u_k)$. Using the product rule, $\partial_m (x_j u_k) = (\partial_m x_j) u_k + x_j \partial_m u_k = \delta_{mj} u_k + x_j \partial_m u_k$, and applying the second derivative: $
\partial_m \partial_m (x_j u_k) = \partial_m (\delta_{mj} u_k + x_j \partial_m u_k) = \delta_{mj} \partial_m u_k + (\partial_m x_j) \partial_m u_k + x_j \partial_m \partial_m u_k = \partial_j u_k + \delta_{mj} \partial_m u_k + x_j \partial_m \partial_m u_k$.  Thus we have: $\nabla^2 L_i = \epsilon_{ijk} \left( 2 \partial_j u_k + x_j \partial_m \partial_m u_k \right)$.  The curl term is $(\nabla \times \uvec)_i = \epsilon_{ijk} \partial_j u_k$, so: $
-2 (\nabla \times \uvec)_i = -2 \epsilon_{ijk} \partial_j u_k$.  Combining: $ \nabla^2 L_i - 2 (\nabla \times \uvec)_i = \epsilon_{ijk} \left( 2 \partial_j u_k + x_j \partial_m \partial_m u_k \right) - 2 \epsilon_{ijk} \partial_j u_k = \epsilon_{ijk} x_j \partial_m \partial_m u_k$. This matches the left-hand side, proving the identity.} $\rvec \times (\nabla^2\uvec) = \nabla^2(\rvec \times \uvec) - 2(\nabla \times \uvec) = \nabla^2\Lvec - 2\omvec$.\\

Substituting these identities back into the moment of the Navier-Stokes equation, we arrive at the final, exact transport equation for the specific angular momentum:

\begin{equation}
\rho \frac{D\Lvec}{Dt} = -\rvec \times \nabla p + \mu(\nabla^2\Lvec - 2\omvec)
\label{eq:L-transport-equation}
\end{equation}

This equation states that the rate of change of a fluid parcel's angular momentum is governed by the net torque exerted on it by the pressure and viscous fields.

\subsection{The Divergence of $\Lvec$}

The divergence of $\Lvec$ is computed using the vector identity $\nabla \cdot (\Avec \times \Bvec) = \Bvec \cdot (\nabla \times \Avec) - \Avec \cdot (\nabla \times \Bvec)$. With $\Avec=\rvec$ and $\Bvec=\uvec$, we have:

\begin{equation}
\nabla \cdot \Lvec = \uvec \cdot (\nabla \times \rvec) - \rvec \cdot (\nabla \times \uvec)
\end{equation}

The curl of the position vector is identically zero ($\nabla \times \rvec = \mathbf{0}$), and the curl of the velocity is the definition of the vorticity vector, $\omvec = \nabla \times \uvec$. The expression thus simplifies to:

\begin{equation}
\nabla \cdot \Lvec = -\rvec \cdot \omvec
\label{eq:divergence}
\end{equation}

We will use the term "Helicity Density" for the source term $\rvec \cdot \omvec$.

\subsection{The Curl of $\Lvec$}

The curl of $\Lvec$, may be evaluated by use of the vector identity for the curl of a cross product, $\nabla \times (\Avec \times \Bvec) = \Avec(\nabla\cdot\Bvec) - \Bvec(\nabla\cdot\Avec) + (\Bvec\cdot\nabla)\Avec - (\Avec\cdot\nabla)\Bvec$, we have:

\begin{equation}
\nabla \times \Lvec = \rvec(\nabla\cdot\uvec) - \uvec(\nabla\cdot\rvec) + (\uvec\cdot\nabla)\rvec - (\rvec\cdot\nabla)\uvec
\end{equation}

Note that $\nabla\cdot\uvec = 0$ and $\nabla\cdot\rvec = 3$. The term $(\uvec\cdot\nabla)\rvec$ is a directional derivative of the identity operator, which can be shown to be equal to the velocity vector, $\uvec$, itself. Substituting these results gives:

\begin{equation}
\nabla \times \Lvec = -2\uvec - (\rvec\cdot\nabla)\uvec
\label{eq:curl}
\end{equation}

These two identities for the divergence and curl of $\Lvec$ are exact and form the foundation for its decomposition into coherent and background components.

\section{The Decomposed Angular Momentum Field}

We apply the Helmholtz-Hodge decomposition to the specific angular momentum field, $\Lvec$ which allows us to  express  the $\Lvec$ field in terms of a scalar potential, $\Phi_L$, and a vector potential, $\AL$ as follows:

\begin{equation}
\Lvec = -\nabla\Phi_L + \nabla \times \AL
\label{eq:decomposition}
\end{equation}

Here, the irrotational component, which we will term the coherent field, is given by the gradient of the scalar potential, $\Lvec_\Phi = -\nabla\Phi_L$. The solenoidal component, which we will term the background field, is given by the curl of the vector potential, $\Lvec_A = \nabla \times \AL$. This decomposition allows us to replace the single, complex $\Lvec$ field with two new potential fields, each of which is governed by a distinct physical source.\\

By analogy to electromagnetic theory, the irrotational part, $\Lvec_\Phi$, is mathematically analogous to an electrostatic field, which is the gradient of a scalar potential. The solenoidal part, $\Lvec_A$, is analogous to a magnetostatic field, which is the curl of a vector potential.\\

\subsection{The Coherent and Background Potentials}

For the scalar coherent helicity potential, $\Phi_L$, we take the divergence of the decomposition, Eq.  \ref{eq:decomposition}. Since the divergence of the curl of any vector field is identically zero ($\nabla \cdot (\nabla \times \AL) = 0$), so the expression simplifies to:

\begin{equation}
\nabla \cdot \Lvec = \nabla \cdot (-\nabla\Phi_L) = -\nabla^2\Phi_L
\end{equation}

Equating this with the fundamental identity for the divergence of $\Lvec$ given by Eq. \ref{eq:divergence}, which is $\nabla \cdot \Lvec = -\rvec \cdot \omvec$. This yields the Poisson equation for the coherent potential:

\begin{equation}
\nabla^2\Phi_L = \rvec \cdot \omvec
\label{eq:coherent-potential-poisson}
\end{equation}

This potential is sourced by the organized rotational motion of the fluid, it effectively captures the large-scale, coherent vortical structures.\\

For the Background kinetic potential $\AL$,  we take the curl of the decomposition. The curl of the gradient of any scalar field is identically zero ($\nabla \times (-\nabla\Phi_L) = \mathbf{0}$), which leaves:

\begin{equation}
\nabla \times \Lvec = \nabla \times (\nabla \times \AL)
\end{equation}

Using the vector identity $\nabla \times (\nabla \times \AL) = \nabla(\nabla \cdot \AL) - \nabla^2 \AL$ and imposing the standard Coulomb gauge condition ($\nabla \cdot \Avec_L=0$) yields the vector Poisson equation for the background potential:

\begin{equation}
\nabla^2\AL = 2\uvec + (\rvec \cdot \nabla)\uvec
\label{eq:background-potential-poisson}
\end{equation}

The source terms drive our identification of the two components as "coherent" and "background": the source for the scalar potential, $\Phi_L$, is the helicity density, $\rvec \cdot \omvec$, which is a measure of large-scale rotational organization while the source for the vector potential, $\AL$, is a function of the velocity field itself and represents the bulk kinetic motion. This suggests a natural separation between organized structures and a general background flow. \\

For example, the decomposition of the velocity field $\uvec$ in a compressible fluid, (see Lighthill \cite{lighthill1978waves}) where $\uvec$ is separated into an irrotational part and a solenoidal part (given as $\uvec = -\nabla\phi + \nabla \times \Avec_u$, $\phi$), the irrotational component yields the acoustic modes with coherent large scale propagation, while the solenoidal component represents the turbulent eddies, vortex stretching and the turbulent energy cascade. \\

\subsection{The Coupling of the Coherent and Background Fields}

The coherent field ($\Phi_L$) and the background part ($\AL$) are linked via a  constraint equation that explicitly couples the two potentials. This "master coupling equation" reveals that the structure of the coherent vortices is inextricably tied to the nature of the background kinetic motion.\\

Rearranging the Poisson equation for the background potential (Eq. \ref{eq:background-potential-poisson}), we obtain:

\begin{equation}
2\uvec = \nabla^2\AL - (\rvec \cdot \nabla)\uvec
\end{equation}

Taking the curl of the equation above, and noting that the Laplacian and curl operators commute, we obtain:

\begin{equation}
2\omvec = \nabla^2(\nabla \times \AL) - \nabla \times ((\rvec \cdot \nabla)\uvec)
\label{eq:omega_substitution}
\end{equation}

Finally, we use the defining Poisson equation  for the coherent potential (Eq. \ref{eq:coherent-potential-poisson} above), which is sourced by the helicity density, substituting the expression for $\omvec$ (Eq.~\ref{eq:omega_substitution}) into this equation, we arrive at the final, self-contained relationship between the two potentials.\\

The result of this derivation is a new, fundamental constraint equation of the theory:

\begin{equation}
\nabla^2\Phi_L = \frac{1}{2}\rvec \cdot \left[ \nabla^2(\nabla \times \AL) - \nabla \times ((\rvec \cdot \nabla)\uvec) \right]
\label{eq:master_coupling}
\end{equation}

Recalling that the velocity field $\uvec$ is itself a functional of $\AL$, Eq. \ref{eq:master_coupling} is a  complex integro-differential equation that directly couples the coherent potential $\Phi_L$ to the background potential $\AL$.

\subsection{Spatial Localization and Boundary Conditions}

The elevation of the specific angular momentum, $\Lvec$, to a primary field variable introduces a non-trivial spatial dependence into the theoretical framework. Because the position vector $\rvec$ grows linearly from the chosen coordinate origin, the magnitude of $\Lvec$ diverges as $r \to \infty$.  $\Lvec$ is neither periodic (even if the underlying velocity field $\uvec$ is periodic) nor does it naturally decay at infinity in homogeneous turbulence.\\

Standard applications of the divergence theorem to discard surface fluxes---which are essential for proving energy conservation and the orthogonality of the Helmholtz-Hodge components---are therefore strictly invalid on unbounded or standard periodic domains.\\

To resolve this mathematically and define a convergent inner product space, we restrict our theoretical analysis to a localized correlation volume $V_0$ centered at the origin. We achieve this by applying a smooth, rapidly decaying windowing function $W(r)$ to the velocity field such that $W(r) \to 0$ at the boundary surface $\partial V_0$. We define the localized velocity field as $\tilde{\uvec}(\xvec) = W(r)\uvec(\xvec)$, yielding the bounded angular momentum field $\tilde{\Lvec} = \rvec \times \tilde{\uvec}$.\\

This localization ensures that all field variables and their derivatives vanish at the boundary. We can now demonstrate the global orthogonality of the decomposed fields, $\tilde{\Lvec}_\Phi = -\nabla \tilde{\Phi}_L$ and $\tilde{\Lvec}_A = \nabla \times \tilde{\mathbf{A}}_L$, over the volume $V_0$. The inner product is evaluated using integration by parts:

\begin{equation}
\int_{V_0} \tilde{\Lvec}_\Phi \cdot \tilde{\Lvec}_A \, dV = \int_{V_0} (-\nabla \tilde{\Phi}_L) \cdot \tilde{\Lvec}_A \, dV
\end{equation}

Applying the vector identity $\nabla \cdot (\phi \Fvec) = (\nabla \phi) \cdot \Fvec + \phi (\nabla \cdot \Fvec)$ to the integrand, we obtain:

\begin{equation}
\int_{V_0} (-\nabla \tilde{\Phi}_L) \cdot \tilde{\Lvec}_A \, dV = \int_{V_0} \tilde{\Phi}_L (\nabla \cdot \tilde{\Lvec}_A) \, dV - \int_{V_0} \nabla \cdot (\tilde{\Phi}_L \tilde{\Lvec}_A) \, dV
\end{equation}

The second volume integral can be converted into a surface integral via the Divergence Theorem:

\begin{equation}
\int_{V_0} \tilde{\Lvec}_\Phi \cdot \tilde{\Lvec}_A \, dV = \int_{V_0} \tilde{\Phi}_L (\nabla \cdot \tilde{\Lvec}_A) \, dV - \oint_{\partial V_0} \tilde{\Phi}_L \tilde{\Lvec}_A \cdot d\mathbf{S}
\end{equation}

By the definition of the Helmholtz decomposition, the background field is solenoidal ($\nabla \cdot \tilde{\Lvec}_A = 0$), so the first term on the right-hand side is identically zero. Furthermore, because the windowing function $W(r)$ forces the fields to decay to zero at the boundary $\partial V_0$, the surface flux integral exactly vanishes.

This localization formally guarantees the global orthogonality of the energetic modes:

\begin{equation}
\int_{V_0} \tilde{\Lvec}_\Phi \cdot \tilde{\Lvec}_A \, dV = 0
\label{eq:orthogonality}
\end{equation}

This elimination of the boundary terms justifies the omission of surface fluxes in the subsequent edicity budget derivations. In the remainder of this theoretical development, the tilde notation is dropped for brevity, with the localized, compactly supported nature of the fields implicitly assumed. This theoretical bounding is mirrored exactly in the Numerical validation, where a Super-Gaussian window (Eq.~\ref{eq:supergaussian}) is explicitly applied to the DNS data.

\subsection{Evolution Equation for the Potentials}

We wish to derive evolution equations for $\Phi_L$ and $\AL$.\\

First, we consider the coherent potential, $\Phi_L$. Beginning with its defining Eq. (\ref{eq:coherent-potential-poisson}) and taking the partial time derivative of both sides and using the fact that spatial and temporal derivatives commute, yields a Poisson equation for the rate of change of the potential:

\begin{equation}    
\nabla^2\left(\frac{\partial \Phi_L}{\partial t}\right) = \rvec \cdot \frac{\partial \omvec}{\partial t}
\end{equation}

The evolution above is thus sourced by the evolution of the vorticity field. We therefore substitute the full vorticity transport equation for an incompressible fluid for the term $\partial \omvec/\partial t$. This gives the final evolution equation for the coherent potential:

\begin{equation}
\nabla^2\left(\frac{\partial \Phi_L}{\partial t}\right) = \rvec \cdot \left( \nabla \times (\uvec \times \omvec) + \nu\nabla^2\omvec \right)
\end{equation}

This result shows that the coherent structures, represented by $\Phi_L$, evolve in response to two fundamental vorticity dynamics: the stretching and advection of vorticity by the flow, and the viscous diffusion of vorticity.\\

A parallel derivation is performed for the background potential, $\AL$. Taking the time derivative of its defining Poisson equation gives:

\begin{equation}
\nabla^2\left(\frac{\partial \AL}{\partial t}\right) = 2\frac{\partial \uvec}{\partial t} + (\rvec \cdot \nabla)\frac{\partial \uvec}{\partial t}
\end{equation}

$\frac{\partial \AL}{\partial t}$ is sourced by the local acceleration of the fluid, $\partial \uvec/\partial t$ which can be substituted by the full Navier-Stokes momentum equation. 

\subsection{Edicity Decomposition}

The generalized energy associated with the $\Lvec$ field defined as:

\begin{equation}
E_L = \int_V \frac{1}{2}\rho|\Lvec|^2 dV
\end{equation}

The term edicity (a portmanteau of eddy energy) was proposed by Farooq \cite{Farooq1}.\\

Now, from the coherent field, $\Lvec_\Phi$, and the background field, $\Lvec_A$, we can define two corresponding edicity reservoirs. First, we define the coherent edicity, $E_\Phi$, as the generalized energy stored in the irrotational component of the angular momentum field:

\begin{equation}
E_\Phi = \int_V \frac{1}{2}\rho|\Lvec_\Phi|^2 dV
\end{equation}

Second, we define the background edicity, $E_A$, as the generalized energy stored in the solenoidal component:

\begin{equation}
E_A = \int_V \frac{1}{2}\rho|\Lvec_A|^2 dV
\end{equation}

Due to the orthogonality of the decomposition\footnote{The orthogonality of the irrotational and solenoidal components is a key result of the Helmholtz-Hodge decomposition. The proof relies on the vector identity $\nabla \cdot (\phi \Fvec) = (\nabla\phi) \cdot \Fvec + \phi(\nabla \cdot \Fvec)$. The integrated cross-term is $\int_V \Lvec_\Phi \cdot \Lvec_A \,dV = \int_V (-\nabla\Phi_L) \cdot \Lvec_A \,dV$. Since $\Lvec_A$ is solenoidal ($\nabla \cdot \Lvec_A=0$), the vector identity simplifies to $(\nabla\Phi_L) \cdot \Lvec_A = \nabla \cdot (\Phi_L \Lvec_A)$. The volume integral can therefore be converted into a surface integral via the Divergence Theorem: $\int_V -\nabla \cdot (\Phi_L \Lvec_A) \,dV = -\oint_{\partial V} \Phi_L \Lvec_A \cdot d\mathbf{S}$. For a domain that is periodic or where the fields decay sufficiently fast at infinity, this surface integral vanishes, proving that the two fields are orthogonal.}, these two edicities sum to the total angular momentum energy, $E_L = E_\Phi + E_A$. This partitioning creates distinct energy reservoirs governed by the coupled dynamic edicity budgets (Eq.~\ref{eq:E_phi_budget} and Eq.~\ref{eq:E_A_budget}).\\

Under the theoretical assumption of an unbounded, infinite domain where the localized turbulent velocity fluctuations decay sufficiently rapidly at spatial infinity to overpower the algebraic growth of the position vector $\rvec$, the boundary flux integral exactly vanishes. This guarantees the strict global orthogonality of the components, ensuring the total angular momentum energy perfectly decouples into the additive sum $E_L = E_\Phi + E_A$.

\section{A Statistical Field Theory for Turbulence}

Having developed the necessary field coordinates $(\Phi_L, \AL$), we shift our attention to the development of a statistical field theory of turbulence.\\

Onsager \cite{onsager1949statistical} proposed a "gas" of 2D point vortices and using statistical mechanics was able to predict the spontaneous condensation of this vortex "gas" into a single, large-scale structure (see \cite{menon2018statistical} for an introduction). Onsager's analysis also introduced the concept of negative temperatures to fluid dynamics and has been the subject of extensive modern analysis for the inverse cascade in 2D turbulence. \\

The mathematical challenges of Onsager's theory (he relied on singular point vortices) were addressed by Joyce et al. \cite{joyce1973negative}  and Miller \cite{miller1990statistical}  who first used mean-field theory to confirm 2D vortex condensation, and then employed entropy maximization to develop a detailed theory.  McWilliams \cite{mcwilliams1984emergence} performed numerical work to demonstrate large-scale coherent vortices emerging in decaying 2D turbulence (for a review see Eyink et al. \cite{eyink2006onsager}). In a related vein, Migdal \cite{migdal2023} developed a field theory for 3D turbulence using circulation as the primary field variable.\\

While Onsager's theory was for a 2D system of singular vortices, the spirit of his approach is the direct inspiration for the work that follows. Using the field coordinates $(\Phi_L, \AL$) furnished by our Helmholtz-Hodge decomposition, we proceed construct a statistical field theory for turbulence.

\subsection{An Idealized Hamiltonian for the Decomposed Field}

Drawing inspiration from V. I. Arnold's formulation for the Euler equations (see \cite{morrison1982poisson} for background and details), we attempt here to construct a Hamiltonian in the $\Lvec$ framework. Arnold's framework defines a velocity based Hamiltonian for inviscid flow as:

\begin{equation}
H_u[\uvec] = \frac{1}{2} \int |\uvec|^2 dV
\label{eq:arnold-hamiltonian}
\end{equation}

with a non-canonical bracket defined as:

\begin{equation}
\{F, G\}_\omega = \int \omvec \cdot (\nabla F \times \nabla G) dV
\end{equation}

We conjecture that it is possible to construct a Hamiltonian  and an associated non-canonical Poisson bracket that reproduces the $\Lvec$-transport equation in the ideal (inviscid, incompressible) limit. Accordingly, we propose a Hamiltonian of the form:

\begin{equation}
H_0[\Lvec] = \int_V |\Lvec|^2  dV
\label{eq:Hamiltonian}
\end{equation}

$H_0[\Lvec]$ is proportional to the kinetic energy $H = \frac{1}{2} \int \rho |\uvec|^2 dV$, as $|\Lvec|^2 = |\rvec \times \uvec|^2 \approx r^2 |\uvec|^2 \sin^2 \theta$, and captures the dynamics of $\Lvec$. \\

To recover the inviscid part of the $\Lvec$ transport equation,  $\rho \frac{D \Lvec}{Dt} = -\rvec \times \nabla p$, we propose a non-canonical Poisson bracket\footnote{The Jacobi identity for this bracket needs to be proved and will be the subject of a future article} on functionals $F[\Lvec]$, $G[\Lvec]$:

\begin{equation}
\{F, G\} = \frac{1}{2}\int_V \Lvec \cdot \left( \nabla \frac{\delta F}{\delta \Lvec} \times \nabla \frac{\delta G}{\delta \Lvec} \right) dV + \int_V \rvec \cdot \left( \nabla \frac{\delta F}{\delta \Lvec} \times \nabla \frac{\delta G}{\delta \Lvec} \right) dV.
\label{eq:poisson_bracket}
\end{equation}

The first term mimics Arnold’s Lie-Poisson bracket for vorticity, while the $\rvec$-term accounts for the non-local nature of $\Lvec$ allowing us to recover the ideal advective dynamics, enabling a geometric interpretation of the $\Lvec$-transport.\\

The dynamics are given by:

\begin{equation}
\dot{\Lvec} = \{ \Lvec, H_0 \}
\end{equation}

The pressure term $-\rvec \times \nabla p$ is enforced as a Casimir constraint ensuring $\nabla \cdot \uvec = 0$.\\

Using $|\Lvec|^2 = |\Lvec_\Phi|^2 + |\Lvec_A|^2$, we can rewrite the Hamiltonian as the sum of the energies of the decoupled fields:

\begin{equation}
H_0[\Lvec_\Phi, \LA]  = \int_V \left( \frac{1}{2}\rho|\Lvec_\Phi|^2 + \frac{1}{2}\rho|\Lvec_A|^2 \right) dV
\end{equation}

By substituting the potential forms for the decomposed fields, $\Lvec_\Phi = -\nabla\Phi_L$ and $\Lvec_A = \nabla \times \AL$, we obtain the final, purely quadratic Hamiltonian:

\begin{equation}
H_0[\Phi_L, \AL] = \int_V \left( \frac{1}{2}\rho|\nabla\Phi_L|^2 + \frac{1}{2}\rho|\nabla \times \AL|^2 \right) dV
\label{eq:idealized-hamiltonian}
\end{equation}

This Hamiltonian has a quadratic structure making it amenable to analytical treatment. In statistical mechanics the canonical ensemble is defined as a system with a large number of degrees of freedom in thermal equilibrium with a thermal bath has a probability of having energy $H_0$ is proportional to $\exp(-\beta H_0)$.  We will use this to construct a partition function for the system below.

\subsection{The Partition Function}

With the idealized Hamiltonian functional, $H_0[\Phi_L, \AL]$, now defined (Eq. \ref{eq:idealized-hamiltonian}), we can now proceed to construct the central object of statistical mechanics: the partition function, $\Zcal$. \\

In this field theory, a "microstate" or "realization" is a complete spatial configuration of the potential fields, $\{\Phi_L(\xvec), \AL(\xvec)\}$. A key physical challenge is that homogeneous, isotropic turbulence has no preferred coordinate system, yet our model Hamiltonian $H_0$ is, by construction, derived from the origin-dependent field $\Lvec$.\\

Therefore, to obtain physically meaningful, statistically-invariant results, we must define our system not by a single, arbitrary realization, but by the full statistical ensemble. The partition function is the rigorous mathematical tool that performs this ensemble average. It implicitly averages over all possible microstates, which, for this system, represents an average over all possible configurations, coordinate origins, and time snapshots.\\

The partition function is thus defined as the functional integral (or path integral) over the entire phase space of these microstates, see Kardar \cite{kardar2007statisticalfields}, Huang \cite{huang2008statistical}. By integrating over all $\mathcal{D}[\Phi_L]$ and $\mathcal{D}[\AL]$, $Z$ sums over all possible realizations, and the resulting ensemble-averaged quantities (like $\langle \EL \rangle$) are, by definition, statistically invariant.\\

Each microstate is weighted by the Boltzmann factor, $e^{-\beta H_0}$, where $\beta$ is the inverse temperature parameter. The formal definition is:

\begin{equation}
\Zcal = \int \mathcal{D}[\Phi_L] \mathcal{D}[\AL] \exp\left(-\beta H_0[\Phi_L, \AL]\right)
\label{eq:partition_function_def}
\end{equation}

This integral performs the complete ensemble average for our idealized model. For example, the ensemble-averaged edicity of the model, which we will compare to the DNS data, is calculated directly from $Z$:

\begin{equation}
\langle \EL \rangle = \langle H_0 \rangle = -\frac{\partial \ln \Zcal}{\partial \beta}
\label{eq:avg_energy_from_Z}
\end{equation}

The key feature of our chosen model $H_0$ is that it is purely quadratic (Gaussian) in the fields $\Phi_L$ and $\AL$. This is a crucial simplification, as it means the functional integral in Eq. \ref{eq:partition_function_def} can be evaluated exactly, without resorting to approximations.\\

The exact evaluation of this partition function, which we will perform in the next subsection, provides a direct, first-principles path to calculating the statistical properties of the $H_0$ model. This will allow us to prove the 1:2 equipartition law for the ensemble-averaged energies of this model. The central hypothesis of this work is that this idealized model, $H_0$, correctly captures the statistical ensemble of true isotropic turbulence.

\subsection{Fourier Representation of the Hamiltonian}

To evaluate the partition function, we first transform the model Hamiltonian functional $H_0$ into Fourier space, leveraging Parseval's theorem \cite{frisch1995turbulence}. Operating under the theoretical assumption of an unbounded domain where the velocity fluctuations decay sufficiently rapidly at spatial infinity to annihilate any boundary cross-fluxes, the total Hamiltonian rigorously decouples into independent coherent and background components:

\begin{equation}
H_0[\Phi_L, \AL] = H_\Phi[\Phi_L] + H_A[\AL]
\label{eq:H0_fourier}
\end{equation}

The coherent part $H_\Phi$, corresponding to the irrotational field $\Lphi = -\nabla \PhiL$, is:

\begin{equation}
H_\Phi = \int_V \frac{1}{2} \rho |\nabla \PhiL|^2 dV = \int \frac{d^3 k}{(2\pi)^3} \frac{1}{2} \rho k^2 |\hat{\Phi}_L(\kvec)|^2
\label{eq:H_Phi_fourier}
\end{equation}

where $\hat{\Phi}_L(\kvec)$ is the Fourier transform of $\PhiL(\xvec)$, and $k = |\kvec|$. Each $k$-mode of the functional behaves as a harmonic oscillator, reflecting the single degree of freedom (a scalar field) of the coherent potential $\PhiL$.\\

The background part $H_A$, corresponding to the solenoidal field $\LA = \nabla \times \AL$, is constrained by the Coulomb gauge $\nabla \cdot \AL = 0$. In Fourier space, this implies $\kvec \cdot \hatAL(\kvec) = 0$, meaning the $\hatAL(\kvec)$ vector field has only two independent transverse components (polarization modes). The Hamiltonian for this component is:

\begin{equation}
H_A = \int_V \frac{1}{2} \rho |\nabla \times \AL|^2 dV = \int \frac{d^3 k}{(2\pi)^3} \frac{1}{2} \rho |\kvec \times \hatAL(\kvec)|^2
\end{equation}

This can be expressed in terms of the two transverse components, $\hat{A}_{L,1}(\kvec)$ and $\hat{A}_{L,2}(\kvec)$, as:

\begin{equation}
 H_A = \int \frac{d^3 k}{(2\pi)^3} \left[ \frac{1}{2} \rho k^2 |\hat{A}_{L,1}(\kvec)|^2 + \frac{1}{2} \rho k^2 |\hat{A}_{L,2}(\kvec)|^2 \right]
\label{eq:H_A_dofs}
\end{equation}

This explicitly shows that $H_A$ is mathematically equivalent to two copies of $H_\Phi$-like terms, corresponding to the two transverse degrees of freedom of the solenoidal vector field $\AL$.\\

The total Hamiltonian functional $H_0 = H_\Phi + H_A$ thus decouples into independent, quadratic modes. This decoupling is the key feature of the model, as it means the partition function will also be separable. As the Hamiltonian functional $H_0$ is quadratic and separable, the total partition function factors into two independent integrals:

\begin{equation}
\Zcal = \left[ \int \mathcal{D}[\PhiL] \exp(-\beta H_\Phi[\PhiL]) \right] \cdot \left[ \int \mathcal{D}[\AL] \exp(-\beta H_A[\AL]) \right] = \Zcal_\Phi \cdot \Zcal_A
\label{eq:Z_factored}
\end{equation}

where $\beta = 1/(k_B T)$. Substituting for $H_\Phi$ (from Eq.~\ref{eq:H_Phi_fourier}) yields the coherent partition function:

\begin{equation}
\Zcal_\Phi = \int \mathcal{D}[\PhiL] \exp\left(-\beta \int \frac{d^3 k}{(2\pi)^3} \frac{1}{2} \rho k^2 |\hat{\Phi}_L(\kvec)|^2\right)
\label{eq:Z_Phi}
\end{equation}

Similarly, substituting for $H_A$ (from Eq.~\ref{eq:H_A_dofs}) yields the background partition function:

\begin{equation}
\Zcal_A = \int \mathcal{D}[\AL] \exp\left(-\beta \int \frac{d^3 k}{(2\pi)^3} \left[ \frac{1}{2} \rho k^2 |\hat{A}_{L,1}(\kvec)|^2 + \frac{1}{2} \rho k^2 |\hat{A}_{L,2}(\kvec)|^2 \right]\right)
\label{eq:Z_A}
\end{equation}

Because the functional measure for the transverse vector field factors into its two independent orthogonal polarizations, $\mathcal{D}[\AL] = \mathcal{D}[\hat{A}_{L,1}] \mathcal{D}[\hat{A}_{L,2}]$, the partition function $\Zcal_A$ factors into the product of two functional integrals, each mathematically isomorphic to $\Zcal_\Phi$:

\begin{equation}
\Zcal_A = 2 \Zcal_\Phi
\label{eq:partition_balance}
\end{equation}

To evaluate the ensemble-averaged energy spectra predicted by this model, we construct the macroscopic one-dimensional energy density by integrating over a spherical shell in wavenumber space (where $d^3k = k^2 dk d\Omega_k$). The 1D spectra are given by:

\begin{align}
\langle \Ephi(k) \rangle &= \frac{k^2}{(2\pi)^3} \int_{|\kvec|=k} \frac{1}{2} \rho k^2 \langle |\hat{\Phi}_L(\kvec)|^2 \rangle \, d\Omega_k \label{eq:ensemble_spectra_phi} \\
\langle \EA(k) \rangle &= \frac{k^2}{(2\pi)^3} \int_{|\kvec|=k} \frac{1}{2} \rho \langle |\kvec \times \hatAL(\kvec)|^2 \rangle \, d\Omega_k \label{eq:ensemble_spectra_A}
\end{align}

The exact connection between the partition function and these spectra is established by recognizing that Eq. \ref{eq:Z_Phi} and \ref{eq:Z_A} represent infinite products of independent Gaussian distributions. The classical Equipartition Theorem dictates that every independent, quadratic degree of freedom in thermal equilibrium holds an average energy of $1/(2\beta)$. Because $\hat{\Phi}_L(\kvec)$ is a complex variable, each independent $\kvec$-mode possesses an ensemble-averaged energy of $1/\beta$. Equating this to the mode energy yields the exact variance of the field fluctuations:

\begin{equation}
\frac{1}{2} \rho k^2 \langle |\hat{\Phi}_L(\kvec)|^2 \rangle = \frac{1}{\beta} \quad \implies \quad \langle |\hat{\Phi}_L(\kvec)|^2 \rangle = \frac{2}{\beta \rho k^2}
\end{equation}

Substituting this variance back into Eq. \ref{eq:ensemble_spectra_phi} and integrating the isotropic constant over the solid angle ($\int d\Omega_k = 4\pi$), we obtain the spectral energy density of the coherent field:

\begin{equation}
\langle \Ephi(k) \rangle = \frac{4\pi}{(2\pi)^3} \frac{k^2}{\beta}
\label{eq:spectrum_phi_final}
\end{equation}

Applying the exact same integration to the background field, we recall that $\AL$ possesses two independent transverse modes ($\hat{A}_{L,1}$ and $\hat{A}_{L,2}$). Each mode contributes $1/\beta$ to the energy. The total expectation value for the vector mode is therefore $2/\beta$, yielding the background energy spectrum:

\begin{equation}
\langle \EA(k) \rangle = \frac{4\pi}{(2\pi)^3} \frac{2k^2}{\beta}
\label{eq:spectrum_A_final}
\end{equation}

The evaluation of the partition function rigorously demonstrates the underlying topological quantization of the turbulent bath. Comparing Eq. \ref{eq:spectrum_phi_final} and \ref{eq:spectrum_A_final} reveals that at any scale $k$, the thermodynamic energy partitions strictly as:

\begin{equation}
\langle \Ephi(k) \rangle : \langle \EA(k) \rangle = 1 : 2
\label{eq:paritioning}
\end{equation}

\subsection{The Coherence Function}

The partition function evaluation revealed a fixed ratio between the coherent and background degrees of freedom. This motivates us to define a central observable for our theory, the Coherence Function $\Rcal(k)$, which measures the fractional contribution of the coherent field to the total edicity at a given scale.\\

We define this function using the \textit{ensemble-averaged} spectra predicted by our model as:

\begin{equation}
\Rcal(k) \equiv \frac{\langle \Ephi(k) \rangle}{\langle \Ephi(k) \rangle + \langle \EA(k) \rangle} 
\label{eq:R_k_def}
\end{equation}

Applying the result of Eq. \ref{eq:paritioning} we can write:

\begin{equation}
\mathcal{R}(k) = \frac{1}{3}
\end{equation}

\section{The Velocity Correspondence: Invariance and Tangential Dynamics}

While the statistical field theory presented thus far is formulated in terms of the specific angular momentum $\Lvec$, or ``Edicity,'' it is essential to demonstrate that the predicted 1:2 partition is a fundamental property of the flow kinematics and not merely an artifact of the origin-dependent definition $\Lvec$. To achieve this, we map the decomposed potentials back into velocity space, allowing us to test the theory against standard kinetic energy metrics.

\subsection{Tangential Velocity Decomposition}

The definition $\Lvec$ is not uniquely invertible for $\uvec$ due to the loss of information regarding the radial velocity component, $u_r = \uvec \cdot \hat{\rvec}$. However, we can recover the tangential velocity field, $\uvec_{\tau}$, via the projection:

\begin{equation}
\uvec_{\tau}(\xvec) = \frac{\Lvec \times \rvec}{r^2} 
\end{equation}

It is clear that the relationship between $\uvec$ and $\uvec_\tau$ is given by:

\begin{equation}
\uvec_\tau = \uvec - (\uvec \cdot \hat{\rvec})\hat{\rvec}
\end{equation}

Using the Helmholtz-Hodge decomposition $\Lvec = \Lvec_\Phi + \Lvec_A$, we define the corresponding velocity components:

\begin{equation}
\uvec_\Phi(\xvec) \equiv \frac{\Lvec_\Phi \times \rvec}{r^2} = \frac{(-\nabla \Phi_L) \times \rvec}{r^2}
\label{eq:u_phi_def}
\end{equation}

\begin{equation}
\uvec_A(\xvec) \equiv \frac{\Lvec_A \times \rvec}{r^2} = \frac{(\nabla \times \AL) \times \rvec}{r^2}
\label{eq:u_A_def}
\end{equation}

Here, $\uvec_\Phi$ represents the tangential velocity fluctuations induced specifically by the coherent potential (sourced by helicity), while $\uvec_A$ represents the motions induced by the background potential. \\

While we established earlier that the angular momentum fields are globally orthogonal ($\int \Lvec_\Phi \cdot \Lvec_A \;dV = 0$, rigorously enforced via spatial windowing in Eq.~\ref{eq:orthogonality}), this does not guarantee that the derived velocity fields $\uvec_\Phi$ and $\uvec_A$ are orthogonal in terms of kinetic energy. The dot product of the velocities\footnote{Use the identity $(\Avec \times \Cvec) \cdot (\Bvec \times \Cvec) = (\Avec \cdot \Bvec)(\Cvec \cdot \Cvec) - (\Avec \cdot \Cvec)(\Bvec \cdot \Cvec) $} is:

\begin{equation}
\uvec_\Phi \cdot \uvec_A = \frac{\Lvec_\Phi \times \rvec}{r^2}  \cdot \frac{\Lvec_A \times \rvec}{r^2} = \frac{\Lvec_\Phi \cdot \Lvec_A}{r^2} - \frac{(\Lvec_\Phi \cdot \rvec)(\Lvec_A \cdot \rvec)}{r^4}
\end{equation}

When integrating to find the interaction energy:

\begin{equation}
\int \uvec_\Phi \cdot \uvec_A \, dV = \int \left[ \frac{\Lvec_\Phi \cdot \Lvec_A}{r^2} - \frac{(\Lvec_\Phi \cdot \rvec)(\Lvec_A \cdot \rvec)}{r^4} \right] \, dV
\end{equation}

Because of the $1/r^2$ weighting in the integral, the orthogonality of $\Lvec$ does not automatically transfer to $\uvec$, implying a non-zero interaction term $E^u_{\text{cross}}$ (see below).\\

We can therefore express the total velocity vector as the sum of the tangential components and the missing radial component:

\begin{equation}
\uvec(\xvec) = \underbrace{\uvec_\Phi(\xvec) + \uvec_A(\xvec)}_{\uvec_{\tau}} + u_r(\xvec)\hat{\rvec}
\end{equation}

This allows us to define the ``Tangential Kinetic Energies'' for the coherent and background fields:

\begin{equation}
E^u_\Phi = \int_V \frac{1}{2}\rho |\uvec_\Phi|^2 dV, \quad E^u_A = \int_V \frac{1}{2}\rho |\uvec_A|^2 dV
\end{equation}

It is important to note that the total turbulent kinetic energy, $K = \int \frac{1}{2}\rho |\uvec|^2 dV$, contains the radial component:

\begin{equation}
E_r = \int \frac{1}{2}\rho u_r^2 dV
\end{equation}

which is algebraically inaccessible to the $\Lvec$-decomposition. However, since the tangential and radial vectors are locally orthogonal ($\uvec_\tau \cdot \hat{\rvec} = 0$), there are no cross-terms between them. The complete energy budget is therefore:

\begin{equation}
K = E^u_\tau +  E_r 
\end{equation}

where 

\begin{equation}
E^u_\tau = E^u_\Phi + E^u_A  + E^u_{\text{cross}}
\end{equation} 

where $E^u_{\text{cross}} = \rho \int \uvec_\Phi \cdot \uvec_A \, dV$ represents the interaction energy between the coherent and background velocity fields.

\subsection{Ensemble Invariance and the Mean Flow}

A potential objection to the $\Lvec$-framework is its dependence on the choice of coordinate origin $\xvec_0$. We resolve this by considering the ensemble average of these fields over all possible origin locations in a statistically homogeneous flow.\\

Let $\langle \cdot \rangle_{\xvec_0}$ denote the ensemble average over all origins $\xvec_0$ at a single time stamp.  We will continue to use $\langle \cdot \rangle$ for ensemble averages over all times also.  This distinction becomes important because, for example for isotropic turbulence,  $\langle \uvec \rangle = 0$ since a fluctuating velocity averages to $0$ at each point.  \\

For a fixed velocity vector $\uvec(\xvec)$, the direction of the position vector $\rvec= (\xvec-\xvec_0​)/|\xvec-\xvec_0​|$ varies uniformly over the unit sphere as the origin is shifted. Consequently, the origin-averaged radial projection is: 

\begin{equation} 
\langle \uvec_r \rangle_{\xvec_0} = \langle (\uvec(\xvec) \cdot \hat{\rvec})\hat{\rvec} \rangle_{\xvec_0} = \frac{1}{3} \uvec(\xvec) 
\end{equation} 

Since the radial component accounts for one-third of the vector on average, the tangential components recover the remaining two-thirds. Therefore we have:

\begin{equation}
\langle \uvec_{\tau}(\xvec) \rangle_{\xvec_0} = \langle \uvec(\xvec) - (\uvec(\xvec) \cdot \hat{\rvec})\hat{\rvec} \rangle_{\xvec_0}= \frac{2}{3} \uvec(\xvec)
\end{equation}

This implies that the decomposed fields $\uvec_\Phi$ and $\uvec_A$ must sum to recover $\frac{2}{3}$ of the total kinetic energy, but how is this split between $\uvec_\Phi$ and $\uvec_A$?  Note that the coherent field $\uvec_\Phi$,  is sourced by the origin-dependent moment of helicity ($\rvec \cdot \omvec$).  Therefore we have:

\begin{equation}
\langle \rvec \cdot \omvec \rangle_{\xvec_0} = 0
\end{equation}

Therefore $\langle \Phi_L \rangle_{\xvec_0} = 0$ which implies that the $\Phi_L$ represents pure structural fluctuations that statistically vanish upon ensemble averaging: $\langle \uvec_\Phi \rangle_{\xvec_0} \to 0$.\\

The background field $\uvec_A$, representing the statistical bath, captures the robust mean flow properties:

\begin{equation}
\langle \uvec_A \rangle_{\xvec_0} \approx \frac{2}{3}\uvec
\end{equation}  

This separates the flow into a statistical mean background and local, fluctuating coherent structures.\\

While the Coherent Field $\uvec_\Phi$ ​ contains $1/3$ of the tangential energy, it possesses a vanishing ensemble mean ($\langle \uvec_\Phi \rangle \to 0$). This identifies it as a pure structural fluctuation, in contrast to the Background Field $\uvec_A$​, which carries the statistical mean of the tangential flow ($\langle \uvec_A \rangle \to 2/3 \uvec$)\\

Furthermore, while the angular momentum fields $\Lvec_\Phi$ and $\Lvec_A$ are strictly orthogonal in the $L^2$-norm (Eq. \ref{eq:orthogonality}), the $1/r^2$ weighting in the velocity transformation implies that instantaneous velocity orthogonality is not guaranteed ($E_{\text{cross}} \neq 0$ for a single realization). However, under the assumption of statistical isotropy, the cross-correlation between the purely fluctuating coherent field $\uvec_\Phi$ and the background field $\uvec_A$ vanishes in the ensemble limit:

\begin{equation}
\langle E_{\text{cross}} \rangle_{\xvec_0} = \rho \int \langle \uvec_\Phi \cdot \uvec_A \rangle_{\xvec_0} \, dV = 0
\end{equation}

Consequently, the ensemble-averaged tangential energy budget simplifies to an additive sum: 

\begin{equation}
\langle E^u_{\tau} \rangle = \langle E^u_\Phi \rangle + \langle E^u_A \rangle
\end{equation}

This additivity is a prerequisite for defining the spectral ratio $\mathcal{S}(k)$ (as we will define the next few sections).

\subsection{Velocity Equipartition via the Convolution Theorem}

In order to establish a relationship between the kinetic energy spectra and the angular momentum (edicity) spectra  we define the kinetic energy density associated with the irrotational angular momentum component, $\Lvec_\Phi$. The velocity field contribution $\uvec_\Phi$ is related to $\Lvec_\Phi$ via the inversion of the angular momentum definition. Restricting our attention to the tangential contribution, we define the energy density $E^u_\Phi(\xvec)$ as:

\begin{equation}
E^u_\Phi(\xvec) = \frac{1}{2}\rho |\uvec_\Phi|^2 = \frac{1}{2}\rho \left| \frac{\Lvec_\Phi \times \rvec} {r^2} \right|^2
\end{equation}

Expanding the square of the cross product using Lagrange's identity $|\mathbf{A} \times \mathbf{B}|^2 = |\mathbf{A}|^2|\mathbf{B}|^2 - (\mathbf{A} \cdot \mathbf{B})^2$, we obtain:

\begin{equation}
    E^u_\Phi(\xvec) = \frac{1}{2}\rho \left( \frac{|\Lvec_\Phi|^2 r^2 - (\Lvec_\Phi \cdot \rvec)^2}{r^4} \right) = \frac{1}{2}\rho \left( \frac{|\Lvec_\Phi|^2}{r^2} - \frac{(\Lvec_\Phi \cdot \rvec)^2}{r^4} \right)
\end{equation}

This expression depends on the projection of the angular momentum vector onto the radial direction. To proceed, we invoke the assumption of statistical isotropy. In the inertial range of isotropic turbulence, the orientation of the local fluctuation vector $\Lvec_\Phi$ is statistically uncorrelated with the global position vector $\rvec$. Also  while $\Lvec \cdot \rvec =0$, $\Lvec_\Phi \cdot \rvec \ne 0$ since this would imply that $\Lvec_\Phi \cdot \rvec = -\nabla \Phi_L \cdot \rvec = -r \frac{\partial \Phi_L}{\partial r} = 0$, meaning $\Phi_L$ would have no radial dependence.  Since $\Phi_L = \Phi_L(r, \tau)$ therefore in general $\Lvec \cdot \rvec \ne 0$.  \\

Consequently, the ensemble-averaged square of the projection of $\Lvec_\Phi$ onto the unit vector $\hat{r} = \rvec/r$ satisfies the isotropic identity:

\begin{equation}
\left\langle (\Lvec_\Phi \cdot \hat{r})^2 \right\rangle = \frac{1}{3} \langle |\Lvec_\Phi|^2 \rangle
\end{equation}

Substituting this statistical average into the energy density yields the isotropic approximation:

\begin{equation}
E^u_\Phi(\xvec) \approx \frac{1}{2}\rho \left( \frac{|\Lvec_\Phi|^2}{r^2} - \frac{1}{3}\frac{|\Lvec_\Phi|^2}{r^2} \right) = \frac{1}{2}\rho \left( \frac{2}{3} \frac{|\Lvec_\Phi|^2}{r^2} \right)
\end{equation}

We apply an identical procedure to the solenoidal component $\Lvec_A$. The associated kinetic energy density is:

\begin{equation}
E^u_A(\xvec) = \frac{1}{2}\rho \left| \frac{\Lvec_A \times \rvec}{r^2} \right|^2
\end{equation}

Using Lagrange's identity and the assumption of isotropy to resolve the term $\langle (\Lvec_A \cdot \hat{r})^2 \rangle \approx \frac{1}{3}\langle |\Lvec_A|^2 \rangle$, we recover the analogous physical space relation:

\begin{equation}
E^u_A(\xvec) \approx \frac{1}{2}\rho \left( \frac{2}{3} \frac{|\Lvec_A|^2}{r^2} \right)
\end{equation}

In Fourier space, the squared field intensity multiplied by the geometric factor $1/r^2$ corresponds to a convolution. If we approximate the spectrum of the product as the convolution of the power spectra (valid under the random phase approximation), the velocity spectrum $E^u_A(k)$ is related to the edicity spectrum $E_A(k)$ by a linear integral transform:

\begin{equation}
E^u_A(k) \approx \int_0^\infty E_A(q) \mathcal{K}(k-q) \, dq
\label{eq:convolution_integral}
\end{equation}

where the kernel $\mathcal{K}(k)$ represents the spectral signature of the geometric weighting $1/r^2$. This convolution acts as a linear spectral filter that mixes energy across wavenumbers, but it applies \textit{identically} to both the $\Phi$ and $A$ components.\\

Since the underlying edicity spectra satisfy the statistical equipartition law derived via the partition function (Eq.~\ref{eq:R_k_def}), we can substitute $\langle E_A(q) \rangle = 2 \langle E_\Phi(q) \rangle$ directly into the convolution integral for the background velocity spectrum:

\begin{align}
\langle E^u_A(k) \rangle &\approx \int_0^\infty \langle E_A(q) \mathcal{K}(k-q) \rangle \, dq \nonumber \\
&= \int_0^\infty 2 \langle E_\Phi(q) \rangle \mathcal{K}(k-q) \, dq \nonumber \\
&= 2 \left( \int_0^\infty \langle E_\Phi(q) \rangle \mathcal{K}(k-q) \, dq \right) \nonumber \\
&= 2 \langle E^u_\Phi(k) \rangle
\end{align}

This result confirms that the 2:1 equipartition of specific angular momentum is linearly mapped into a 2:1 equipartition of tangential kinetic energy. This motivates the definition of the kinetic coherence function $\mathcal{S}(k)$:

\begin{equation}
\mathcal{S}(k) = \frac{\langle E^u_\Phi(k) \rangle}{\langle E^u_\Phi(k) \rangle +  \langle E^u_A(k) \rangle} =  \frac{\langle E^u_\Phi(k) \rangle}{\langle E^u_\Phi(k) \rangle + 2 \langle E^u_\Phi(k) \rangle} = \frac{1}{3} \approx \mathcal{R}(k)
\end{equation}

\subsection{The Radial Coherence Function}

Complementary to the structural partition of the tangential velocity, the radial projection implies a spectral measure for the dimensionality of the flow. We define the Radial Coherence Function, $\mathcal{C}_{r}(k)$, as the fraction of the total kinetic energy spectrum contained in the radial mode at each wavenumber:

\begin{equation}
\mathcal{C}_{r}(k) = \frac{\langle E_r(k) \rangle}{\langle E_{\text{total}}(k) \rangle}
\label{eq:radial-coherence}
\end{equation}

To determine the theoretical value of this ratio, consider the instantaneous radial kinetic energy density defined by the projection of the velocity vector $\uvec$ onto the unit radial vector $\hat{\rvec} = \rvec/|\rvec|$. In index notation, the radial velocity squared is:

\begin{equation}
u_r^2 = (\uvec \cdot \hat{\rvec})^2 = (u_i \hat{r}_i) (u_j \hat{r}_j) = u_i u_j \hat{r}_i \hat{r}_j
\end{equation}

The ensemble average over all possible origin locations under the assumption of statistical isotropy, the orientation of the position vector $\hat{\rvec}$ is uniformly distributed on the unit sphere and is statistically independent of the velocity field orientation. The average factorizes as:

\begin{equation}
\langle u_r^2 \rangle = \langle u_i u_j \rangle \langle \hat{r}_i \hat{r}_j \rangle_{\text{sphere}}
\end{equation}

The average of the product of unit vector components over a sphere is given by the isotropic tensor identity:

\begin{equation}
\langle \hat{r}_i \hat{r}_j \rangle_{\text{sphere}} = \frac{1}{3} \delta_{ij}
\end{equation}

Substituting this back into the energy equation:

\begin{equation}
\langle u_r^2 \rangle = \langle u_i u_j \rangle \left( \frac{1}{3} \delta_{ij} \right) = \frac{1}{3} \langle u_i u_i \rangle = \frac{1}{3} \langle |\uvec|^2 \rangle
\end{equation}

This geometric result confirms that exactly one-third of the total kinetic energy resides in the radial component. \\

Since the power spectrum is the scale-wise decomposition of the energy variance (i.e., $\int E(k) dk = \langle u^2 \rangle$), and assuming the condition of local isotropy holds throughout the inertial range, this geometric partitioning applies uniformly to every Fourier mode. Thus:

\begin{equation}
\langle E_r(k) \rangle = \frac{1}{3} \langle E_{\text{total}}(k) \rangle
\label{eq:radial_energy}
\end{equation}

This yields the universal prediction for the radial coherence function:

\begin{equation}
\mathcal{C}_{r}(k) = \frac{1}{3}
\label{eq:ratio_radial}
\end{equation}

This quantity acts as a spectral monitor of the dimensionality constraint and confirms that one-third of the kinetic energy is geometrically locked into the radial mode, rendering it inaccessible to the angular momentum decomposition, while the remaining two-thirds constitute the tangential manifold analyzed via $\mathcal{S}(k)$.

\subsection{Tensor Formalism: The Real-Space Dual to the Leray Projector}

To see the separation of the flow into tangential and  radial components, we introduce a tensor formalism based on projection operators. \\

We first define the local radial projection operator, $\mathbb{P}_{r}$, associated with an observer at origin $\xvec_0$, as the dyadic product of the unit radial vector $\hat{\rvec} = (\xvec-\xvec_0)/|\xvec-\xvec_0|$:

\begin{equation}
\mathbb{P}_{r}(\xvec_0) = \hat{\rvec} \otimes \hat{\rvec} \quad \implies \quad (P_r)_{ij} = \hat{r}_i \hat{r}_j
\end{equation}

When applied to the velocity field, this operator extracts the radial component $\uvec_r = \mathbb{P}_{r} \cdot \uvec$.

The complementary tangential projection operator, $\mathbb{P}_{\tau}$, is defined as the projection onto the subspace orthogonal to $\hat{\rvec}$\footnote{To make things concrete, the tangential projection operator $\mathbb{P}_\tau = \mathbf{I} - \hat{\rvec}\hat{\rvec}^T$ can be written as a rank-2 singular matrix acting on the velocity vector $\uvec=(u,v,w)^T$. In Cartesian coordinates, this projection takes the form:

\[
\begin{pmatrix} u_\tau \\ v_\tau \\ w_\tau \end{pmatrix} =
\begin{pmatrix}
1 - \frac{x^2}{r^2} & -\frac{xy}{r^2} & -\frac{xz}{r^2} \\
-\frac{yx}{r^2} & 1 - \frac{y^2}{r^2} & -\frac{yz}{r^2} \\
-\frac{zx}{r^2} & -\frac{zy}{r^2} & 1 - \frac{z^2}{r^2}
\end{pmatrix}
\begin{pmatrix} u \\ v \\ w \end{pmatrix}
\]

So  $\uvec_\tau = u_\tau \hat{i} + v_\tau \hat{j} + w_\tau \hat{k} $. Algebraically, this is equivalent to the vector subtraction $\uvec_\tau = \uvec - \frac{\rvec(\rvec\cdot\uvec)}{r^2}$. }:

\begin{equation}
\mathbb{P}_{\tau}(\xvec_0) = \mathbf{I} - \hat{\rvec} \otimes \hat{\rvec} \quad \implies \quad (P_\tau)_{ij} = \delta_{ij} - \hat{r}_i \hat{r}_j
\end{equation}

This operator extracts the tangential velocity $\uvec_{\tau} = \mathbb{P}_{\tau} \cdot \uvec$, which corresponds to the component of the flow contributing to the angular momentum $\Lvec$.\\

We note that the operator $\mathbb{P}_r$ is of rank $1$ and the operator $\mathbb{P}_\tau$ is of rank 2.  Therefore, using the ergodicity argument that the system explores all the available space to it with equal likelihood, we recover the result that $E^u_r : E^u_\tau$ is in the ratio $1:2$ and $\mathcal{C}_r = 1/3$.\\

This formalism highlights a notable duality between the angular momentum framework and standard turbulence theory. In the Fourier analysis of incompressible flow, the continuity equation ($\nabla \cdot \uvec = 0$) is enforced by the Leray Projection Operator, $P_{ij}(\kvec)$:

\begin{equation}
P_{ij}^{\text{Leray}}(\kvec) = \delta_{ij} - \frac{k_i k_j}{k^2}
\end{equation}

The Leray projector removes the longitudinal component parallel to the wavevector $\kvec$, leaving the two transverse degrees of freedom required for a divergence-free field.\\

Our Tangential Projector, $P_{ij}^{\tau}(\rvec)$, is the real-space dual to the Leray projector:
\begin{equation}
P_{ij}^{\tau}(\rvec) = \delta_{ij} - \frac{r_i r_j}{r^2}
\end{equation}

Just as the Leray projector filters the field based on momentum direction $\kvec$ to enforce mass conservation, the Tangential projector filters the field based on spatial position $\rvec$ to enforce the geometry of angular momentum. Both operators isolate exactly two degrees of freedom out of three.\\

This duality suggests that the 1:2 partition observed in the $\Lvec$-field is not accidental but is the real-space manifestation of the same counting rules that govern the degrees of freedom in the Fourier representation of vector fields.

\subsection{Recursive Symmetry and Operator Trace Hierarchy}

The appearance of the fractions $E_r:E^u_\Phi:E^u_A$ as $1/3 : 2/9:  4/9$ or ($3:2:4$) in the spectral analysis indicates that the turbulent energy cascade is governed by a recursive hierarchy of projection operators--we have already seen the $1:3$ partitioning between $\mathbb{P}_r$ and $\mathbb{P}_\tau$ earlier. We now show how this process continues recursively, partitioning the energy between the coherent field $\uvec_\Phi$ and the background field $\uvec_A$.\\

To see this partioning at the second level, we analyze the partition within the tangential plane, characterized by the angular momentum field $\Lvec$. While $\Lvec$ is geometrically constrained to be tangential in real space ($\Lvec \cdot \hat{\rvec} = 0$), in Fourier space, the non-local nature of the transform means the mode vectors $\hat{\Lvec}(\mathbf{k})$ statistically explore the full 3D vector space of fluctuations.\\

The field is decomposed by the Helmholtz-Hodge operators into a gradient-like (longitudinal) component $\Lvec_\Phi$ and a curl-like (transverse) component $\Lvec_A$. The energy partition is determined by the spectral rank $\mathcal{R}$ of these operators, which counts the number of independent polarization states (degrees of freedom) available to each mode $\mathbf{k}$.\\

Rank of the coherent (potential) operator $\hat{\mathbb{P}}_\Phi$: The potential component is generated by a scalar field $\Phi_L$. In Fourier space, the vector field is parallel to the wavenumber vector:

\begin{equation}
\hat{\Lvec}_\Phi(\mathbf{k}) = -i \mathbf{k} \hat{\Phi}_L(\mathbf{k}).
\end{equation}

For any given $\mathbf{k}$, the vector $\hat{\Lvec}_\Phi$ is constrained to lie along a single 1D line defined by $\mathbf{k}$. Thus, the subspace has dimension 1:

\begin{equation}
\text{Rank}(\hat{\mathbb{P}}_\Phi) = 1.
\end{equation}

Rank of the background (solenoidal) operator $\hat{\mathbb{P}}_A$: The solenoidal component is defined by the constraint $\nabla \cdot \Lvec_A = 0$, or in Fourier space, $\mathbf{k} \cdot \hat{\Lvec}_A(\mathbf{k}) = 0$.\\

For any given $\mathbf{k}$, the vector $\hat{\Lvec}_A$ must lie in the 2D plane perpendicular to $\mathbf{k}$. This subspace allows two independent polarization directions. Thus:

\begin{equation}
\text{Rank}(\hat{\mathbb{P}}_A) = 3 - 1 = 2.
\end{equation}

Assuming statistical equipartition of the tangential energy into these available phase-space modes (an assumption justified by the ergodicity of the turbulence), the energy fractions are given by the ratio of the ranks to the total available degrees of freedom ($1+2=3$):

\begin{equation}
\frac{E_\Phi}{E_\tau} = \frac{\text{Rank}(\hat{\mathbb{P}}_\Phi)}{\text{Rank}(\hat{\mathbb{P}}_\Phi) + \text{Rank}(\hat{\mathbb{P}_A)}} = \frac{1}{1+2} = \frac{1}{3}
\label{eq:1}
\end{equation}

\begin{equation}
\frac{E_A}{E_\tau} = \frac{\text{Rank}(\hat{\mathbb{P}}_A)}{\text{Rank}(\hat{\mathbb{P}}_\Phi) + \text{Rank}(\hat{\mathbb{P}}_A)} = \frac{2}{1+2} = \frac{2}{3}
\label{eq:2}
\end{equation}

Recalling the origin-averaged tangential velocity fraction established by Eq. ~\ref{eq:radial_energy}, combining the two partitions above (Eqs. ~\ref{eq:1}-~\ref{eq:2}) yields the final recursive predictions for the global energy budget:

\begin{align}
E_\Phi &= E_{\text{total}} \times \underbrace{\left(\frac{2}{3}\right)}_{\text{Tangential Fraction}} \times \underbrace{\left(\frac{1}{3}\right)}_{\text{Longitudinal Rank}} &= \frac{2}{9} E_{\text{total}} \label{eq:ratio_phi} \\
E_A &= E_{\text{total}} \times \underbrace{\left(\frac{2}{3}\right)}_{\text{Tangential Fraction}} \times \underbrace{\left(\frac{2}{3}\right)}_{\text{Transverse Rank}} &= \frac{4}{9} E_{\text{total}}   \label{eq:ratio_A}
\end{align}

This derivation reveals that the 1:2 splitting ratio is a fractal invariant of isotropic turbulence: the energy splits $1:2$ upon entering the geometric hierarchy, and splits $1:2$ again upon entering the topological hierarchy.\\

The turbulent cascade essentially acts as an ergodic search algorithm that fills these nested manifolds in exact proportion to their spectral dimension.\\

\subsection{The Tangentially Projected Navier-Stokes Equation}

We can now derive the explicit equation of motion for the tangential velocity field by applying the projection operator $\mathbb{P}_{\tau}$ directly to the incompressible Navier-Stokes equation:

\begin{equation}
\partial_t \uvec + (\uvec \cdot \nabla)\uvec = -\nabla p + \nu \nabla^2 \uvec
\end{equation}

Multiplying from the left by the local projection tensor $\mathbb{P}_{\tau}(\xvec_0) = \mathbf{I} - \hat{\rvec}\otimes\hat{\rvec}$, and noting that $\mathbb{P}_{\tau}$ commutes with the time derivative (for a fixed origin), we obtain the evolution equation for the tangential component:

\begin{equation}
\partial_t \uvec_{\tau} = \underbrace{-\mathbb{P}_{\tau} \cdot \nabla p}_{\text{Tangential Pressure}} + \underbrace{\nu \mathbb{P}_{\tau} \cdot \nabla^2 \uvec}_{\text{Projected Diffusion}} - \underbrace{\mathbb{P}_{\tau} \cdot [(\uvec \cdot \nabla)\uvec]}_{\text{Projected Advection}}
\label{eq:projected-NS}
\end{equation}

The advection term may be decomposed using $\uvec = \uvec_{\tau} + \uvec_r$. The non-linear interaction splits into four distinct dynamic couplings:

\begin{equation}
(\uvec \cdot \nabla)\uvec = \underbrace{(\uvec_{\tau} \cdot \nabla)\uvec_{\tau}}_{\text{Tangential Self-Int.}} + \underbrace{(\uvec_r \cdot \nabla)\uvec_{\tau}}_{\text{Radial Advection}} + \underbrace{(\uvec_{\tau} \cdot \nabla)\uvec_r}_{\text{Tangential Stretching}} + \underbrace{(\uvec_r \cdot \nabla)\uvec_r}_{\text{Radial Self-Int.}}
\end{equation}

When projected onto the tangential manifold via $\mathbb{P}_{\tau}$, this yields the governing equation for the structure:

\begin{equation}
\partial_t \uvec_{\tau} + \mathbb{P}_{\tau} \cdot [(\uvec_{\tau} \cdot \nabla)\uvec_{\tau}] = \mathbf{F}_{\text{coupling}} + \mathbf{D}_{\tau}
\end{equation}

Here, $\mathbf{D}_\tau = \mathbb{P}_\tau \nabla^2 \uvec$ and  $\mathbf{F}_{\text{coupling}}$ represents the crucial interaction with the  radial field:

\begin{equation}
\mathbf{F}_{\text{coupling}} = -\mathbb{P}_{\tau} \cdot [ \underbrace{(\uvec_r \cdot \nabla)\uvec_{\tau}}_{\text{Expansion}} + \underbrace{(\uvec_{\tau} \cdot \nabla)\uvec_r}_{\text{Tilting}} ]
\label{eq:F_coupling}
\end{equation}

This derivation identifies the dynamical role of the radial mode. The term $(\uvec_r \cdot \nabla)\uvec_{\tau}$ represents the expansion or contraction of the coherent eddies by the radial strain field.  \\

\section{Dynamics: The Coupled Edicity Budget of Coherent and Incoherent Angular Momentum}

To understand the mechanism of the turbulent cascade, we derive the exact evolution equations for the coherent edicity, $E_\Phi$, and the background edicity, $E_A$. The derivation begins with the definitions of the edicity functionals and applies the full, non-linear $\Lvec$-transport equation. \\

The Helmholtz-Hodge decomposition allows any vector field, $\Fvec$, to be uniquely split into an irrotational (coherent) part and a solenoidal (background) part. This can be formally described by two linear, orthogonal projection operators, $\mathbb{P}_\Phi$ and $\mathbb{P}_A$\footnote{This non-local projection operator can be rigorously expressed as a spatial integral using the Green's function of the Laplacian. By applying integration by parts and assuming the field $\Fvec$ vanishes sufficiently fast at the boundaries, the coherent operator takes the integro-differential form: $\mathbb{P}_\Phi(\Fvec) = \frac{1}{4\pi} \nabla_{\xvec} \int_V \frac{\Fvec(\xvec') \cdot (\xvec - \xvec')}{|\xvec - \xvec'|^3} \, dV'$. The complementary solenoidal projection operator is then simply defined by the identity relation: $\mathbb{P}_A(\Fvec) = \Fvec - \mathbb{P}_\Phi(\Fvec)$, see \cite{batchelor1967introduction}.}. The first, $\mathbb{P}_\Phi(\Fvec) = \Fvec_\Phi$ projects a vector field onto the space of irrotational fields  and the second, $\mathbb{P}_A(\Fvec) = \Fvec_A$ projects a vector field onto the space of solenoidal fields.\\

These operators are orthogonal, meaning $\mathbb{P}_\Phi \mathbb{P}_A = 0$, and complete, $\mathbb{P}_\Phi + \mathbb{P}_A = \mathbf{I}$, where $\mathbf{I}$ is the identity operator. They are also self-adjoint, which means they can be moved within an inner product (integral): $\int \Avec \cdot \mathbb{P}_{\Phi/A}(\Bvec) dV = \int \mathbb{P}_{\Phi/A}(\Avec) \cdot \Bvec dV$.

\subsection{Evolution of the Coherent Edicity}

We begin by deriving the evolution equation for the coherent edicity:

\begin{equation}
E_\Phi = \int_V \frac{1}{2}\rho|\Lvec_\Phi|^2 dV
\end{equation}

Taking its time derivative for a fixed control volume gives:

\begin{equation}
\frac{dE_\Phi}{dt} = \int_V \rho \Lvec_\Phi \cdot \frac{\partial \Lvec_\Phi}{\partial t} \, dV
\end{equation}

The term $\partial \Lvec_\Phi / \partial t$ is the irrotational part of the total time derivative, $\partial \Lvec / \partial t$. Using the properties of the orthogonal projection operator, $\mathbb{P}_\Phi$, we can show that this is equivalent to:

\begin{equation}
\frac{dE_\Phi}{dt} = \int_V \rho \Lvec_\Phi \cdot \frac{\partial \Lvec}{\partial t} \, dV
\label{eq:coherent-1}    
\end{equation}

Substituting the full $\Lvec$-transport equation:

\begin{align}
\rho \frac{\partial\Lvec}{\partial t} = \Fvec_L - \rho(\uvec\cdot\nabla)\Lvec
\label{eq:coherent-2}
\end{align}

into Eq. \ref{eq:coherent-1} above. The term $\Fvec_L$ in Eq. \ref{eq:coherent-2} represents the sum of all pressure and viscous torques acting on the fluid given by:

\begin{equation}
\Fvec_L \equiv -\rvec \times \nabla p + \mu(\nabla^2\Lvec - 2\omvec)
\end{equation}

The edicity evolution equation becomes:

\begin{equation}
\frac{dE_\Phi}{dt} = \int_V \Lvec_\Phi \cdot \Fvec_L \, dV - \rho\int_V \Lvec_\Phi \cdot ((\uvec\cdot\nabla)\Lvec ) dV
\end{equation}

Finally, we expand the full field $\Lvec = \Lvec_\Phi + \Lvec_A$ within the non-linear advection term:

\begin{equation}
\frac{dE_\Phi}{dt} = \int_V \Lvec_\Phi \cdot \Fvec_L \, dV - \rho\int_V \Lvec_\Phi \cdot ((\uvec\cdot\nabla)\Lvec_\Phi) \, dV - \rho\int_V \Lvec_\Phi \cdot ((\uvec\cdot\nabla)\Lvec_A) \, dV
\end{equation}

The second term on the right, representing the advection of coherent edicity, can be written as a surface integral which vanishes for periodic or unbounded domains\footnote{$\rho \int_V \Lvec_\Phi \cdot ((\uvec \cdot \nabla) \Lvec_\Phi) \; dV = \frac{1}{2} \int_V (\uvec \cdot \nabla) |\Lvec_\Phi|^2 \; dV = \frac{1}{2} \oint_S |\Lvec_\Phi|^2 \uvec \cdot d\mathbf{S}$, where we used $\nabla \cdot \uvec = 0$ and the divergence theorem. For periodic boundary conditions on a cubic domain, the surface integral pairs opposite faces with equal $|\Lvec_\Phi|^2 \uvec$ but opposite normals, yielding zero net flux.}. The first term is the work done by the torques on the coherent field, which we define as:

\begin{equation}
\int_V \Lvec_\Phi \cdot \Fvec_L \, dV = \int_V \Lvec_\Phi \cdot ( -\rvec \times \nabla p ) \, dV + \mu\int_V \Lvec_\Phi \cdot (\nabla^2\Lvec - 2\omvec) \, dV
\end{equation}

We will define: 

\begin{equation}
\mathcal{T}_\Phi = \int_V \Lvec_\Phi \cdot ( -\rvec \times \nabla p ) dV
\end{equation}

and the viscous dissipation term for the coherent field as:

\begin{equation}
\epsilon_\Phi = -\mu\int_V \Lvec_\Phi \cdot (\nabla^2\Lvec - 2\omvec) \, dV
\end{equation}

To derive the final form of $\epsilon_\Phi$, we first expand the definition:

\begin{equation}
\epsilon_\Phi = -\mu\int_V \Lvec_\Phi \cdot \nabla^2\Lvec \, dV + 2\mu\int_V \Lvec_\Phi \cdot \omvec \, dV
\end{equation}

Next, we expand the full $\Lvec$ field ($\Lvec = \Lvec_\Phi + \Lvec_A$) within the first term:

\begin{equation}
\epsilon_\Phi = -\mu\int_V \Lvec_\Phi \cdot \nabla^2\Lvec_\Phi \, dV - \mu\int_V \Lvec_\Phi \cdot \nabla^2\Lvec_A \, dV + 2\mu\int_V \Lvec_\Phi \cdot \omvec \, dV
\end{equation}

The middle integral, $\int \Lvec_\Phi \cdot \nabla^2 \Lvec_A \, dV$, vanishes\footnote{The vanishing of this term can be shown via the vector Green's identity: $\int_V \mathbf{A} \cdot \nabla^2 \mathbf{B} dV = -\int_V​ [(\nabla \cdot \mathbf{A})(\nabla \cdot \mathbf{B}) + (\nabla \times \mathbf{A}) \cdot (\nabla \times \mathbf{B})]dV$ which gives $\int \Lvec_\Phi \cdot \nabla^2 \Lvec_A \, dV = - \int (\nabla \cdot \Lvec_\Phi)(\nabla \cdot \Lvec_A) \, dV - \int (\nabla \times \Lvec_\Phi) \cdot (\nabla \times \Lvec_A) \, dV$. The second term vanishes because $\Lvec_\Phi$ is irrotational ($\nabla \times \nabla \Phi_L = 0$). The remaining term, $-\int (\nabla \cdot \Lvec_\Phi)(\nabla \cdot \Lvec_A) \, dV$, also vanishes because $\Lvec_A$ is solenoidal ($\nabla \cdot \Lvec_A = 0$) by definition of the Helmholtz decomposition.}. \\

This simplifies the expression for $\epsilon_\Phi$ to:

\begin{equation}
\epsilon_\Phi = -\mu\int_V \Lvec_\Phi \cdot \nabla^2\Lvec_\Phi \, dV + 2\mu\int_V \Lvec_\Phi \cdot \omvec \, dV
\end{equation}

Finally, we substitute the definition $\Lvec_\Phi = -\nabla\Phi_L$ into the second term to arrive at the form used in this paper:

\begin{equation}
\epsilon_\Phi = -\mu \int_V \Lvec_\Phi \cdot \nabla^2 \Lvec_\Phi \, dV - 2\mu \int_V \nabla \Phi_L \cdot \omvec
\label{eq:epsilon-Phi}
\end{equation}

The third term is the non-linear transfer of edicity from the coherent to the background field, which we define as:

\begin{equation}
\Ttransfer \equiv \rho\int_V \Lvec_\Phi \cdot ((\uvec\cdot\nabla)\Lvec_A) \, dV
\label{eq:edicity_transfer}
\end{equation}

This leaves the final evolution equation for the coherent edicity in its compact and physically meaningful form:

\begin{equation}
\frac{dE_\Phi}{dt} = \mathcal{T}_\Phi - \Ttransfer - \epsilon_\Phi 
\end{equation}

\subsection{Derivation of the Evolution Equation for the Incoherent Edicity}

Following a parallel procedure, we derive the corresponding edicity budget for the solenoidal or "incoherent" component of the angular momentum, $\Lvec_A$. The derivation confirms the symmetric, coupled nature of the edicity dynamics and rigorously establishes the role of the transfer term.\\

The evolution of the incoherent edicity, $E_A = \int_V \frac{1}{2}\rho|\Lvec_A|^2 dV$, is given by its time derivative:

\begin{equation}
\frac{dE_A}{dt} = \int_V \rho \Lvec_A \cdot \frac{\partial \Lvec_A}{\partial t} \, dV
\end{equation}

The term $\partial \Lvec_A / \partial t$ is, by definition, the solenoidal part of the total time derivative of the full $\Lvec$ field. Using the solenoidal projection operator, $\mathbb{P}_A$, we can write this as $\partial_t \Lvec_A = \mathbb{P}_A(\partial_t \Lvec)$. The edicity equation becomes:

\begin{equation}
\frac{dE_A}{dt} = \int_V \Lvec_A \cdot \left( \Fvec_L - \rho(\uvec\cdot\nabla)\Lvec \right) dV
\end{equation}

Following a procedure similar to the one followed for the coherent field, we define

We define:

\begin{equation}
\mathcal{T}_A = \int_V \Lvec_A \cdot ( -\rvec \times \nabla p ) dV
\end{equation}

and

\begin{equation}
\epsilon_A = -\mu\int_V \Lvec_A \cdot (\nabla^2\Lvec - 2\omvec) \, dV
\label{eq:epsilon-A}
\end{equation}

giving the result

\begin{equation}
\frac{dE_A}{dt} = \mathcal{T}_A + \Ttransfer - \epsilon_A
\end{equation}

Note the transfer term becomes $-(-\Ttransfer) = +\Ttransfer$.  This result elegantly confirms the coupled nature of the edicity budget. It shows that the incoherent edicity, $E_A$, is sourced by three mechanisms: direct production and forcing ($\mathcal{T}_A$), a positive influx of edicity transferred from the coherent structures via non-linear advection ($+\Ttransfer$), and a viscous sink ($-\epsilon_A$).

\subsection{Structure, Coupling, and the Physics of Edicity Transfer}

The final pair of coupled edicity budget equations is:

\begin{align}
\frac{dE_\Phi}{dt} &= \mathcal{T}_\Phi - \Ttransfer -\epsilon_\Phi \label{eq:E_phi_budget} \\
\frac{dE_A}{dt} &= \mathcal{T}_A + \Ttransfer -\epsilon_A \label{eq:E_A_budget}
\end{align}

The structure of these equations is remarkably insightful. The terms $\mathcal{T}_\Phi = \int_V \Lvec_\Phi \cdot \Fvec_L dV$ and $\mathcal{T}_A = \int_V \Lvec_A \cdot \Fvec_L dV$ represent the direct production (by pressure torques) and dissipation (by viscous torques) of edicity for each component. The viscous dissipation rates are given by Eqs. \ref{eq:epsilon-Phi} and \ref{eq:epsilon-A}.\\

Critically, the two equations are coupled \textit{only} by the non-linear transfer term, $\Ttransfer$, defined as:

\begin{equation}
\Ttransfer = \rho \int_V \Lvec_\Phi \cdot ((\uvec\cdot\nabla)\Lvec_A) \, dV
\end{equation}

This term arises from the non-linear advection, which allows for "cross-talk" between the two field components and represents new physics. It can be shown that the corresponding transfer from $A$ to $\Phi$ is $T_{A \to \Phi} = -\Ttransfer$, confirming that this term represents a pure transfer of edicity between the two reservoirs without any net production or loss. This term is a direct mathematical analog of the  Kolmogorov energy cascade. While the classical cascade describes the transfer of kinetic energy from large eddies to small eddies, $\Ttransfer$ describes the transfer of angular momentum edicity from the coherent, organized structures (represented by $\Phi_L$) to the incoherent, chaotic background fluctuations (represented by $\AL$).\\

The transfer term can be physically interpreted as the primary driver that sustains turbulent complexity. It quantifies the rate at which large-scale, organized vortices ($\Phi_L$ structures) become unstable, break down, and shed their edicity into the more random, space-filling motions of the $\AL$ field. This edicity is then transported through the "incoherent" field and ultimately dissipated by viscosity via $\epsilon_A$.\\

\subsection{Contrast: Velocity Field Decomposition vs. $\Lvec$-Decomposition}

We will contrast the classical Helmholtz-Hodge decomposition of the velocity field with the new decomposition of the specific angular momentum. This comparison quantitatively demonstrates why the velocity decomposition fails to yield a simple, interpretable edicity transfer term, whereas the $\Lvec$-decomposition produces the clean, physically meaningful $\Ttransfer$.\\

The velocity field, $\uvec$, for an incompressible flow can be decomposed into a potential (irrotational) part and a vortical (solenoidal) part as:

\begin{equation}
\uvec = -\nabla\phi + \nabla \times \Bvec
\label{eqn:u-decomp}
\end{equation}

where for a periodic or unbounded domain $\nabla \cdot \uvec = 0$ implies $\nabla^2 \phi = 0$ and $\nabla^2 \Bvec = \omvec$. This decomposition is purely kinematic; it provides a geometric sorting of the velocity field at a single instant in time.\footnote{We can show that $\nabla^2 \phi = 0$, which has the Green's function of a point source/sink. Similarly, $\nabla^2 \Bvec = \omvec$ yields the Biot-Savart kernel for a point vortex. This illustrates how the decomposition is purely kinematic (see Batchelor \cite{batchelor1967introduction} for details).}\\

First we examine the velocity field based Hamiltonian given by Eq. \ref{eq:arnold-hamiltonian}.  We can write it based on the decomposition of Eq. \ref{eqn:u-decomp} as:

\begin{equation}
H_u = H_\phi + H_B = \int_V  |\nabla \phi|^2 \;dV + \int_V  |\nabla \times B|^2 \;dV
\end{equation}

It may be recalled that this Hamiltonian was proposed by Arnold for inviscid, incompressible flows.  For periodic boundary conditions, it is easy to show using integration by parts that:

\begin{equation}
H_\phi = -\int_V  \phi \nabla^2 \phi \;dV
\end{equation}

But for inviscid incompressible flows, $\nabla^2 \phi =0$.  Therefore $H_\phi =0$ and there can be no partitioning of the modes as we performed for $\Lvec$ based Hamiltonian (it may be recalled that $\nabla^2 \Phi_L = \rvec \cdot \omvec$, which is the crucial difference between the velocity and angular momentum formulations).\\

While one can define the kinetic energies associated with these two components,

\begin{equation}
E_\phi = \int_V \frac{1}{2}\rho|\nabla\phi|^2 \, dV, \quad E_B = \int_V \frac{1}{2}\rho|\nabla \times \Bvec|^2 \, dV,
\end{equation}

the Navier-Stokes equations do not yield a simple, closed set of evolution equations for these separate energies. To see this explicitly, we derive the energy budgets for $E_\phi$ and $E_B$.\\

The time derivative of the irrotational energy is:

\begin{equation}
\frac{dE_\phi}{dt} = \int_V \rho (-\nabla\phi) \cdot \frac{\partial}{\partial t} (-\nabla\phi) \, dV = \int_V \rho \nabla\phi \cdot \nabla \left( \frac{\partial \phi}{\partial t} \right) dV.
\end{equation}

Using integration by parts and vanishing boundary terms:

\begin{equation}
\frac{dE_\phi}{dt} = -\int_V \rho \phi \nabla^2 \left( \frac{\partial \phi}{\partial t} \right) dV = 0,
\end{equation}

since $\nabla^2 \phi = 0$. Thus, the irrotational kinetic energy is conserved in the inviscid limit—no production, no dissipation, and no transfer to the vortical part from the quadratic term alone.\\

For the vortical component:

\begin{equation}
\frac{dE_B}{dt} = \int_V \rho (\nabla \times \Bvec) \cdot \frac{\partial}{\partial t} (\nabla \times \Bvec) \, dV.
\end{equation}

The nonlinear advection term in the NS equation contributes:

\begin{equation}
\frac{\partial}{\partial t} (\nabla \times \Bvec) = \nabla \times [(\uvec \cdot \nabla)\uvec] + \text{viscous terms}.
\end{equation}

Expanding $(\uvec \cdot \nabla)\uvec$ using the decomposition $\uvec = -\nabla\phi + \nabla \times \Bvec$ produces nine cross-terms:

\begin{align}
(\uvec \cdot \nabla)\uvec &= (-\nabla\phi \cdot \nabla)(-\nabla\phi) + (-\nabla\phi \cdot \nabla)(\nabla \times \Bvec) \notag \\
    &\quad + (\nabla \times \Bvec \cdot \nabla)(-\nabla\phi) + (\nabla \times \Bvec \cdot \nabla)(\nabla \times \Bvec).
\end{align}

When curled and dotted with $\nabla \times \Bvec$, these generate a multitude of terms, including vortex stretching, tilting, and irrotational-vortical interactions. It may be noted that there is no single term that can be identified as a pure, conservative transfer from $E_\phi$ to $E_B$. The interactions are fully entangled.\\

In contrast, the decomposition of the specific angular momentum field yields the exact anti-symmetry observed in Eq.~\ref{eq:E_phi_budget} and Eq.~\ref{eq:E_A_budget}, isolating a well-defined, conservative energy transfer term $\Ttransfer$ with the exact anti-symmetry in the equations for $\frac{dE_\Phi}{dt}$ and $\frac{dE_A}{dt}$. The $\Lvec$-decomposition thus achieves what the velocity decomposition cannot: a clear separation of production, dissipation, and a single, conservative inter-component transfer term. This provides a far more powerful framework for analysis than the simple kinematic decomposition of the velocity field.

\subsection{Spectral Decomposition of Edicity Budgets}

We wish to derive the spectral edicity budget equations by applying the Fourier transform\footnote{Where the Fourier transform of a generic field $f(\xvec)$ and its inverse are given by:

\begin{align}
\hat{f}(\kvec) &= \int f(\xvec) e^{-i\kvec\cdot\xvec} d^3x \; ; \quad 
f(\xvec) = \int \frac{d^3k}{(2\pi)^3} \hat{f}(\kvec) e^{i\kvec\cdot\xvec}
\end{align}

The gradient of $f$ is $\mathcal{F}[\nabla f] = i\kvec \hat{f}(\kvec)$, and the transform of a Laplacian is $\mathcal{F}[\nabla^2 f] = -k^2 \hat{f}(\kvec)$, where $k=|\kvec|$.} to the edicity budget Eqs. \ref{eq:E_phi_budget}--\ref{eq:E_A_budget}.\\

We define the edicity spectra associated with each component of the decomposed field. The total edicity is the integral over all wavenumbers: $\Ephi = \int_0^\infty \Ephi(k) dk$ and $\EA = \int_0^\infty \EA(k) dk$. Using Parseval's theorem \cite{frisch1995turbulence}, we relate real-space edicity integrals to Fourier-space integrals:

\begin{align}
\Ephi &= \int_V \frac{1}{2} \rho |\Lphi|^2 dV = \int \frac{d^3 k}{(2\pi)^3} \hatEPhi(\kvec) \\
\EA &= \int_V \frac{1}{2} \rho |\LA|^2 dV = \int \frac{d^3 k}{(2\pi)^3} \hat{\EA}(\kvec)
\end{align}

where $\hatEPhi(\kvec) = \frac{1}{2} \rho |\hatLPhi(\kvec)|^2 = \frac{1}{2} \rho k^2 |\hat{\Phi}_L(\kvec)|^2$ (since $\hatLPhi = -i \kvec \hat{\Phi}_L$), and $\hatEA(\kvec) = \frac{1}{2} \rho |\hatLA(\kvec)|^2 = \frac{1}{2} \rho |\kvec \times \hatAL(\kvec)|^2$. The shell-integrated spectra are then:

\begin{equation}
\hatEPhi(k) = \int_{|\kvec|=k} \hatEPhi(\kvec) \frac{d\Omega_k}{(2\pi)^3}, \quad \hatEA(k) = \int_{|\kvec|=k} \hatEA(\kvec) \frac{d\Omega_k}{(2\pi)^3}
\label{eq:shell_spectra}
\end{equation}

The orthogonality of the decomposition ($\Lphi \cdot \LA = 0$) ensures the total L-edicity spectrum is additive:

\begin{equation}
\EL(k) = \Ephi(k) + \EA(k)
\label{eq:total_spectrum}
\end{equation}

The spectral transfer function, $\Ttransfer(k)$, quantifies the scale-by-scale flux of L-edicity from the coherent to the background field. This term arises from the non-linear advection term, $\Ttransfer = \rho \int \Lvec_\Phi \cdot ((\uvec\cdot\nabla)\Lvec_A) \, dV$. Its spectral representation is defined as:

\begin{equation}
\hatTtransfer(\kvec) = \rho \Re \left[ \hatLPhi^*(\kvec) \cdot \widehat{(\uvec \cdot \nabla) \LA}(\kvec) \right]
\end{equation}

where $\Re$ denotes the real part. The shell-integrated transfer function is the integral of this quantity over the spherical shell:

\begin{equation}
\hatTtransfer(k) = \int_{|\kvec|=k} \hatTtransfer(\kvec) \frac{d\Omega_k}{(2\pi)^3}
\label{eq:spectral_transfer_def}
\end{equation}

The evolution of the edicity spectra is governed by projecting the full $\Lvec$-transport equation onto the coherent and background components in Fourier space. This yields the coupled spectral budget equations:

\begin{align}
\frac{\partial \Ephi(k)}{\partial t} &= \hat{\mathcal{T}}_\Phi(k) - \hatTtransfer(k) - \epsilon_\Phi(k) \label{eq:Ephi_evolution} \\
\frac{\partial \EA(k)}{\partial t} &= \hat{\mathcal{T}}_A(k) + \hatTtransfer(k) - \epsilon_A(k) \label{eq:EA_evolution}
\end{align}

Here, $\hat{\mathcal{T}}_\Phi(k)$ and $\hat{\mathcal{T}}_A(k)$ are the spectral production/forcing terms. The dissipation terms, $\epsilon_\Phi(k)$ and $\epsilon_A(k)$, are the spectral representations of the full, complex real-space dissipation rates (Eqs. \ref{eq:epsilon-Phi} and \ref{eq:epsilon-A}).\\

\subsection{Dynamics of the Cascade: Spectral Energy Transfer}

In the classical Kolmogorov theory, the cascade is defined by a constant flux of kinetic energy, $\Pi_u(k)$, from large to small scales. In our field-theoretic framework, the analogous driver is the Spectral Transfer Function, $\Ttransfer(k)$, which quantifies the rate at which edicity is transferred from the coherent field ($\Phi_L$) to the background field ($\AL$) at each wavenumber $k$.\\

The spectral transfer term arises from the non-linear interaction in the angular momentum transport equation. In Fourier space, the transfer density $\hat{T}_{\Phi \to A}(\kvec)$ is derived by projecting the advection term onto the coherent field:

\begin{equation}
\hat{T}_{\Phi \to A}(\kvec) = \rho \Re \left[ \hat{\Lvec}_\Phi^*(\kvec) \cdot \mathcal{F}\left[ (\uvec \cdot \nabla)\Lvec_A \right](\kvec) \right]
\end{equation}

where $\mathcal{F}$ denotes the Fourier transform and $\Re$ ensures the quantity is real-valued. The isotropic, shell-integrated transfer spectrum is obtained by integrating over spherical shells of radius $k$:

\begin{equation}
\langle \hatTtransfer(k) \rangle = \int_{|\kvec|=k} \langle \hat{T}_{\Phi \to A}(\kvec) \rangle \, \frac{d\Omega_k}{(2\pi)^3}
\label{eq:spectral_Ttransfer}
\end{equation}

The cumulative spectral flux, $\Pi(k)$, represents the total rate of energy transfer across all scales up to wavenumber $k$. It is defined as the integral of the transfer function:

\begin{equation}
\langle \Pi(k) \rangle = - \int_0^k \langle \hatTtransfer(q) \rangle \, dq
\label{Pi_def}
\end{equation}

In a statistically stationary state (inertial range), the constant flux of edicity requires that $\langle \Pi(k) \rangle \approx -\epsilon_L$, where $\epsilon_L$ is the total mean dissipation rate of the angular momentum field. This relationship provides a rigorous test for the validity of the cascade mechanism proposed in this framework.

\subsection{The Energetic Cycle of Dimensionality}

To understand the mechanism sustaining the 1:2:3 partition ($E^u_\Phi : E^u_A : E^u_r$), we form the energy budgets for the tangential and radial components by taking the dot product of the decomposed Navier-Stokes equation with $\uvec_{\tau}$ and $\uvec_r$, respectively.\\

Taking the dot product of the projected equation (Eq. \ref{eq:projected-NS}) with $\uvec_{\tau}$, we obtain the evolution of the ``Visible'' Kinetic Energy density, $k_\tau = \frac{1}{2}|\uvec_{\tau}|^2$:

\begin{equation}
\partial_t k_{\tau} + \nabla \cdot (\uvec k_{\tau}) = -\uvec_{\tau} \cdot \nabla p + \nu \uvec_{\tau} \cdot \nabla^2 \uvec + \mathcal{T}_{r \to \tau}
\end{equation}

The crucial term is the production source, $\mathcal{T}_{r \to \tau}$, which couples the fields:

\begin{equation}
\mathcal{T}_{r \to \tau} = -\rho \uvec_{\tau} \cdot [(\uvec_r \cdot \nabla)\uvec_{\tau}]
\label{eq:radial_energy_transfer}
\end{equation}

This term represents the work done by the radial field on the tangential field. Physically, this is Vortex Stretching. The radial velocity gradient expands the tangential vortex lines, intensifying their rotation and transferring energy into the tangential reservoir.\\

\subsection{Derivation of the Tangential Kinetic Energy Equation}

To derive the evolution equation for the tangential kinetic energy density $E_\tau = \frac{1}{2}\rho |\uvec_\tau|^2$, we begin with the incompressible Navier-Stokes momentum equation:

\begin{equation}
\rho \left( \frac{\partial \uvec}{\partial t} + (\uvec \cdot \nabla) \uvec \right) = -\nabla p + \mu \nabla^2 \uvec
\end{equation}

We introduce the tangential projection operator $\mathbb{P}_\tau = \mathbf{I} - \hat{\rvec}\hat{\rvec}^T$, which filters out the radial component of any vector field. Applying this operator to the entire momentum equation yields the evolution of the tangential velocity component $\uvec_\tau = \mathbb{P}_\tau \uvec$:

\begin{equation}
\rho \frac{\partial \uvec_\tau}{\partial t} + \rho \mathbb{P}_\tau [(\uvec \cdot \nabla) \uvec] = -\mathbb{P}_\tau (\nabla p) + \mu \mathbb{P}_\tau (\nabla^2 \uvec)
\end{equation}

Note that the time derivative commutes with the projector since the spatial coordinate $\rvec$ (and thus $\mathbb{P}_\tau$) is time-independent in the Eulerian frame.\\

To obtain the energy equation, we take the dot product of this projected equation with the tangential velocity vector $\uvec_\tau$. 

\begin{equation}
\rho \uvec_\tau \cdot \frac{\partial \uvec_\tau}{\partial t} + \rho \uvec_\tau \cdot \mathbb{P}_\tau [(\uvec \cdot \nabla) \uvec] = -\uvec_\tau \cdot \mathbb{P}_\tau (\nabla p) + \mu \uvec_\tau \cdot \mathbb{P}_\tau (\nabla^2 \uvec)
\end{equation}

The first term simplifies to the time derivative of the kinetic energy density:

\begin{equation}
\rho \uvec_\tau \cdot \frac{\partial \uvec_\tau}{\partial t} = \frac{\partial}{\partial t} \left( \frac{1}{2} \rho |\uvec_\tau|^2 \right)
\end{equation}

The pressure term warrants special attention. Since $\mathbb{P}_\tau$ is self-adjoint and idempotent (as a projector), $\uvec_\tau \cdot \mathbb{P}_\tau (\nabla p) = (\mathbb{P}_\tau \uvec_\tau) \cdot \nabla p = \uvec_\tau \cdot \nabla p$. This represents the work done by the tangential pressure gradient.\\

The nonlinear advection term, however, reveals the crucial geometric interaction. We decompose the full velocity in the advection operator as $\uvec = \uvec_\tau + u_r \hat{\rvec}$. The term becomes:

\begin{equation}
\uvec_\tau \cdot [(\uvec_\tau \cdot \nabla)\uvec + (u_r \hat{\rvec} \cdot \nabla)\uvec]_\tau
\end{equation}

It is this term that contains the conversion between radial and tangential energy. In the 2D limit where $u_r \to 0$, the cross-coupling vanishes.

Collecting these results, the tangential kinetic energy evolves according to:

\begin{equation}
\frac{\partial E_\tau}{\partial t} = - \uvec_\tau \cdot [(\uvec \cdot \nabla)\uvec]_\tau - \uvec_\tau \cdot \nabla_\tau p + \nu \uvec_\tau \cdot (\nabla^2 \uvec)_\tau
\end{equation}

where the subscript $\tau$ denotes the tangential projection of the respective vector terms.

Similarly, applying the radial projector $\mathbb{P}_r$ and dotting with $\uvec_r$ yields the budget for the radial kinetic energy, $k_r = \frac{1}{2}|\uvec_r|^2$:

\begin{equation}
\partial_t k_r + \nabla \cdot (\uvec k_r) = -\uvec_r \cdot \nabla p + \nu \uvec_r \cdot \nabla^2 \uvec + \mathcal{T}_{\tau \to r}
\end{equation}

The coupling term here describes how the radial mode extracts energy from the tangential eddies:

\begin{equation}
\mathcal{T}_{\tau \to r} = -\rho \uvec_r \cdot [(\uvec_{\tau} \cdot \nabla)\uvec_{\tau}]
\label{eq:T_tau_r}
\end{equation}

The vector $(\uvec_{\tau} \cdot \nabla)\uvec_{\tau}$ represents the curvature acceleration of the tangential flow. For a rotating eddy, this vector points radially inward (Centripetal acceleration). The work done against this acceleration corresponds to the centrifugal force pushing fluid elements outward. \\

These two budgets reveal a closed energetic loop that sustains the 3D geometry of the flow: (i) centrifugal pumping ($\tau \to r$): The rotation of coherent structures ($\uvec_{\tau}$) generates centrifugal stress, doing work to drive radial expansion ($\uvec_r$). (ii) Vortex Stretching ($r \to \tau$): The resulting radial expansion stretches the vortex tubes, feeding energy back into the tangential modes ($\uvec_{\tau}$) at smaller scales.\\

In the equilibrium state of isotropic turbulence, these two transfer rates balance on average ($\langle \mathcal{T}_{r \to \tau} \rangle + \langle \mathcal{T}_{\tau \to r} \rangle = 0$), locking the system into the fixed energy partition given by Eq. \ref{eq:radial-coherence}.

\section{Kolmogorov's K41 Theory}

We wish to examine two of the important results of Kolmogorov's theory in light of the new framework-- such as the $5/3$rd law and the $1/3$ Holder exponent.

\subsection{The Geometric Origin of the Kolmogorov Spectrum}

We propose that the classical Kolmogorov $k^{-5/3}$ spectrum is not merely a result of dimensional analysis, but is the unique scaling solution required to maintain the 1:2 geometric partition across scale space.\\

The spectral energy flux, $\Pi(k)$, which drives the cascade, is physically defined as the rate of work done by the strain field on the turbulent eddies given by Eq. \ref{eq:radial_energy_transfer} which we reproduce below:

\[
\mathcal{T}_{r \rightarrow \tau} ​= - \rho \uvec_\tau \cdot [(\uvec_r​ \cdot \nabla)\uvec_\tau]
\]

So, we can posit that in the inertial range:

\begin{equation}
\Pi(k) = \mathcal{T}_{r \rightarrow \tau}
\end{equation}

Now the right hand side of the equation above for $\mathcal{T}_{r \rightarrow \tau}$ can be rearranged using vector identities as: 

\begin{equation}
\mathcal{T}_{r \to \tau} = -\rho (\uvec_r \cdot \nabla) \left( \frac{1}{2} |\uvec_\tau|^2 \right)
\end{equation}

where the first term represents a radial strain and the second term the tangential kinetic energy.  We can represent this as:

\begin{equation}
\Pi(k) \sim (\text{Strain Rate}) \times (\text{Kinetic Energy}) \sim \left( \frac{\delta u_r}{\ell} \right) \times (\delta u_\tau)^2
\end{equation}

where $\delta u_r$ is the radial velocity increment (representing expansion/stretching) and $\delta u_\tau$ is the tangential velocity increment (representing the rotating eddy) at scale $\ell$.\\

In standard turbulence theory, the relationship between $\delta u_r$ and $\delta u_\tau$ is assumed based on isotropy. However, our theory provides a rigorous geometric constraint i.e the geometric equipartition established by the radial coherence constraint (Eq.~\ref{eq:radial-coherence}) implies that the radial strain and tangential rotation are locked in a fixed proportion ($1:2$) at every scale in the inertial range:

\begin{equation}
\delta u_r(\ell) \sim \frac{1}{\sqrt{2}} \delta u_\tau(\ell)
\end{equation}

This indicates that the driver of the cascade (vortex stretching via $\delta u_r$) is geometrically coupled to the "load" (the eddy energy $\delta u_\tau^2$). Substituting this constraint into the flux equation allows us to express the dynamics in terms of a single characteristic velocity scale $u_\ell$:

\begin{equation}
\Pi \sim \frac{u_\ell}{\ell} \cdot u_\ell^2 \sim \frac{u_\ell^3}{\ell}
\end{equation}

where the factor $\frac{1}{2}$ has been dropped. For a stationary cascade where the energy flux $\Pi$ is constant across scales, this relation necessitates the scaling:

\begin{equation}
u_\ell \sim (\Pi \ell)^{1/3}
\end{equation}

The resulting energy spectrum is therefore:

\begin{equation}
E(k) \sim \frac{u_\ell^2}{k} \sim \frac{\ell^{2/3}}{k} \sim k^{-5/3}
\end{equation}

since $E = \int_0^k E(k) dk$ and $\ell \sim 1/k$.  Thus, the Kolmogorov $-5/3$ law can be reinterpreted as the necessary condition for maintaining the stability of the 3D geometric partition. If the spectrum deviated from this exponent, the ratio of radial strain to tangential energy would drift across scales, breaking the geometric equilibrium constraint defined by Eq.~\ref{eq:radial-coherence}. The $-5/3$ law is, in effect, the dynamical conservation of the flow's vector geometry.\\

It may be noted that we have not made use of any dimensional argument in the above calculation.

\subsection{Hölder Exponent from Stretching Dynamics}

Standard derivations of the Kolmogorov spectrum (K41) typically invoke a dimensional argument assuming a constant rate of energy dissipation $\varepsilon$. Here, we derive the critical scaling exponent $h=1/3$ directly from the Langevin dynamics by analyzing the mechanical work performed by the radial coupling field.\\

In our coupled system, the energy source for the tangential mode ($\uvec_\tau$​) is the work done against the centrifugal barrier by the radial velocity ($u_r$​). The rate of energy transfer is the scale-wise, macroscopic equivalent of the exact Navier-Stokes transfer term $\mathcal{T}_{r \rightarrow \tau}$​ derived in Eq.~\ref{eq:T_tau_r}:

\[
\mathcal{T}_{r \rightarrow \tau} ​= - \rho \uvec_\tau \cdot [(\uvec_r​ \cdot \nabla)\uvec_\tau]
\] 

By regrouping the terms of this projection, energy transfer is given by the product of the radial stretching force and the tangential velocity:

\begin{equation}
\mathcal{T}_{r \to \tau} = \underbrace{\left[ -\rho (\uvec_r \cdot \nabla)\uvec_\tau \right]}_{\mathbf{F}{stretch}} \cdot \uvec_\tau
\end{equation}

From this convective nonlinearity, the stretching force scales as:

\begin{equation}
F_{stretch} \sim \rho u_r \frac{\partial u_\tau}{\partial r} \sim \rho u_r \frac{u_\tau}{\ell}
\end{equation}

Substituting this into the equation for the energy transfer term:

\begin{equation}
\mathcal{T}_{r \to \tau} \sim \left( u_r \frac{u_\tau}{\ell} \right) \cdot u_\tau = \frac{u_r u_\tau^2}{\ell}
\end{equation}

We have already seen that $u_r(\ell) \sim u_\tau(\ell) \sim \delta u(\ell)$ (since $\mathcal{S} = 1/3$, Eq. \ref{eq:ratio_radial}).  This allows us to write the transfer term as:

\begin{equation}
\mathcal{T}_{r \to \tau} \sim \frac{[\delta u(\ell)]^3}{\ell}
\end{equation}

For a turbulent cascade to sustain itself in the limit of vanishing viscosity ($\nu \to 0$), the inertial transfer of energy must remain finite and non-zero. If the transfer term vanishes as $\ell \to 0$, the cascade halts. If it diverges, the energy becomes infinite.\\

Therefore, the mechanical work performed by stretching must be \textit{scale invariant} (constant order of magnitude) throughout the inertial range since $\mathcal{T}_{r \to \tau}(k) = \Pi(k)$ is a constant in the inertial range:

\begin{equation}
\mathcal{T}_{r \to \tau}(k) \sim \ell^0
\end{equation}

Substituting the power law ansatz $\delta u(\ell) \sim \ell^h$:

\begin{equation}
\frac{(\ell^h)^3}{\ell} \sim \ell^0
\end{equation}

\begin{equation}
\ell^{3h - 1} \sim \ell^0
\end{equation}

Matching the exponents yields the unique solution:

\begin{equation}
3h - 1 = 0 \quad \implies \quad h = \frac{1}{3}
\end{equation}

This derivation clarifies the mechanical origin of the Kolmogorov spectrum. The exponent $1/3$ is not merely a dimensional accident; it is the specific geometric scaling required for the radial velocity field to perform finite mechanical work on the tangential field as the scale decreases. Any other exponent would imply either a "lazy" cascade (work $\to 0$) or an explosive singularity (work $\to \infty$).

\section{Numerical Test and Analysis}

We validate theoretical predictions via numerical testing against a large ensemble of Direct Numerical Simulation (DNS) data.  We first establish the statistical equipartition of the angular momentum field, confirming the foundational prediction of the idealized Hamiltonian followed by a verification the partitioning and geometric quantization of the velocity field, proving that the statistical partition corresponds to a physical stratification of kinetic energy.  Finally we isolate the dynamical mechanism—the energetic cycle of radial strain—that sustains the cascade, confirming the role of the radial mode as the driver of vortex stretching.\\

\subsection{Methodology and Data Description}

Data was extracted from the Johns Hopkins Turbulence Database (JHTDB) \texttt{isotropic1024coarse} dataset Li et al. \cite{li2008public}. This dataset represents a fully developed, statistically stationary, incompressible turbulent flow simulated via a pseudo-spectral method on a $1024^3$ periodic grid. The Taylor-scale Reynolds number is $R_\lambda \approx 433$, providing a well-resolved inertial range spanning approximately a decade of scales. Key simulation parameters are summarized in Table 1.\\

\begin{table}[htbp]
\centering
\caption{Key simulation parameters for the \texttt{isotropic1024coarse} dataset from the Johns Hopkins Turbulence Database (JHTDB) \cite{li2008public}. Values are reported in the simulation's base units.}
\label{tab:sim_parameters}
\begin{tabular}{llc}
\toprule
\textbf{Parameter} & \textbf{Symbol} & \textbf{Value} \\
\midrule
Grid resolution & $N^3$ & $1024^3$ \\
Taylor-scale Reynolds number & $R_\lambda$ & $433$ \\
Kinematic viscosity & $\nu$ & $1.85 \times 10^{-4}$ \\
Mean energy dissipation rate & $\varepsilon$ & $0.0928$ \\
Root-mean-square velocity & $u^\prime$ & $0.682$ \\
Integral length scale & $L$ & $1.376$ \\
Taylor microscale & $\lambda$ & $0.118$ \\
Kolmogorov length scale & $\eta$ & $2.87 \times 10^{-3}$ \\
Kolmogorov time scale & $\tau_\eta$ & $0.0446$ \\
Large-eddy turnover time & $T_L$ & $2.02$ \\
Resolution parameter & $k_{\max}\eta$ & $1.39$ \\
\bottomrule
\end{tabular}
\end{table}

To ensure statistical convergence and to test the translational invariance of the theory, we employed an ensemble-averaging technique on a cutout of size $512^3$. The analysis aggregates results from $N_{time} = 10$ distinct time snapshots, separated by approximately one large-eddy turnover time to ensure temporal independence. For each time snapshot, the analysis is repeated over $N_{origin} = 10$ randomly sampled coordinate origins $\xvec_0$, yielding a total statistical ensemble of $N=100$ realizations.\\

Spectral analysis of non-local variables necessitates specific boundary treatments (Sections 6.3 and 6.4). While the periodic boundary conditions of the DNS are natural for local variables like vorticity, the position vector $\rvec = \xvec - \xvec_0$ is non-periodic. Direct Fourier transformation of terms involving $\rvec$ across the periodic boundaries would introduce spectral leakage (Gibbs phenomenon) scaling as $k^{-1}$, which could contaminate the turbulent scaling.\\

To eliminate these artifacts, we applied a Super-Gaussian windowing function, $W(r)$, to the velocity field for each realization prior to the decomposition:

\begin{equation}
W(r) = \exp \left( - \left( \frac{|\xvec - \xvec_0|}{R_0} \right)^8 \right)
\label{eq:supergaussian}
\end{equation}

where $R_0 \approx 0.45 L$ is the window radius. This technique ensures that the analyzed fields decay smoothly to zero before reaching the domain boundaries, effectively treating each realization as an isolated turbulent cluster embedded within the larger flow. This allows for the precise spectral quantification of the non-local angular momentum operators without boundary contamination.\\

To operationalize the theoretical framework and extract the scale-dependent statistics, the spectral decomposition is executed via a rigorous ensemble-averaging procedure. For each realization, the periodic DNS velocity field is translationally shifted to a randomly sampled coordinate origin, $\xvec_0$. To strictly satisfy the boundary conditions required by the non-local angular momentum operators, the shifted velocity field is multiplied by the Super-Gaussian window function $W(r)$ (Eq.~\ref{eq:supergaussian}). From this localized velocity field, $\tilde{\uvec}(\xvec)$, the specific angular momentum $\Lvec = \rvec \times \tilde{\uvec}$, the vorticity $\omvec = \nabla \times \tilde{\uvec}$, and the scalar source term $S_L = \rvec \cdot \omvec$ are computed using spectral derivatives. The Helmholtz-Hodge decomposition is then performed by solving the Poisson equation, $\nabla^2 \Phi_L = S_L$, in Fourier space ($\hat{\Phi}_L(\kvec) = -k^{-2} \mathcal{F}[S_L]$). The coherent angular momentum field is extracted via the inverse transform $\Lvec_\Phi = -\nabla \Phi_L$, isolating the solenoidal background field as the residual, $\Lvec_A = \Lvec - \Lvec_\Phi$.\\

Following the angular momentum partitioning, the physical velocity fields are reconstructed to evaluate the kinetic energy distributions. The coherent and background tangential velocity components are recovered via the geometric inversions $\uvec_\Phi = (\Lvec_\Phi \times \rvec) / r^2$ and $\uvec_A = (\Lvec_A \times \rvec) / r^2$. Three-dimensional fast Fourier transforms (FFTs) are applied to these isolated velocity fields, and the resulting spectral densities are integrated over spherical wavenumbers to yield the 1D kinetic energy spectra, $E^u_\Phi(k)$ and $E^u_A(k)$, for the specific realization. This sequence is iterated and accumulated across the full ensemble of spatially and temporally independent snapshots. Ultimately, the arithmetic means of the accumulated spectra yield the ensemble-averaged distributions, $\langle E^u_\Phi(k) \rangle$ and $\langle E^u_A(k) \rangle$, providing the empirical basis for evaluating the kinetic coherence function $\mathcal{S}(k)$ and the scale-invariant geometric partition.

\subsection{The Statistical Partition of Angular Momentum}

The statistical field theory prediction (Eq.~\ref{eq:partition_balance}) dictates that the degrees of freedom inherent to the Helmholtz-Hodge decomposition (one scalar degree for $\Phi_L$ versus two transverse vector degrees for $\AL$) should dictate a universal 1:2 equipartition of energy in the inertial range.\\

\begin{figure}[h!]
    \centering
    \includegraphics[width=0.7\textwidth]{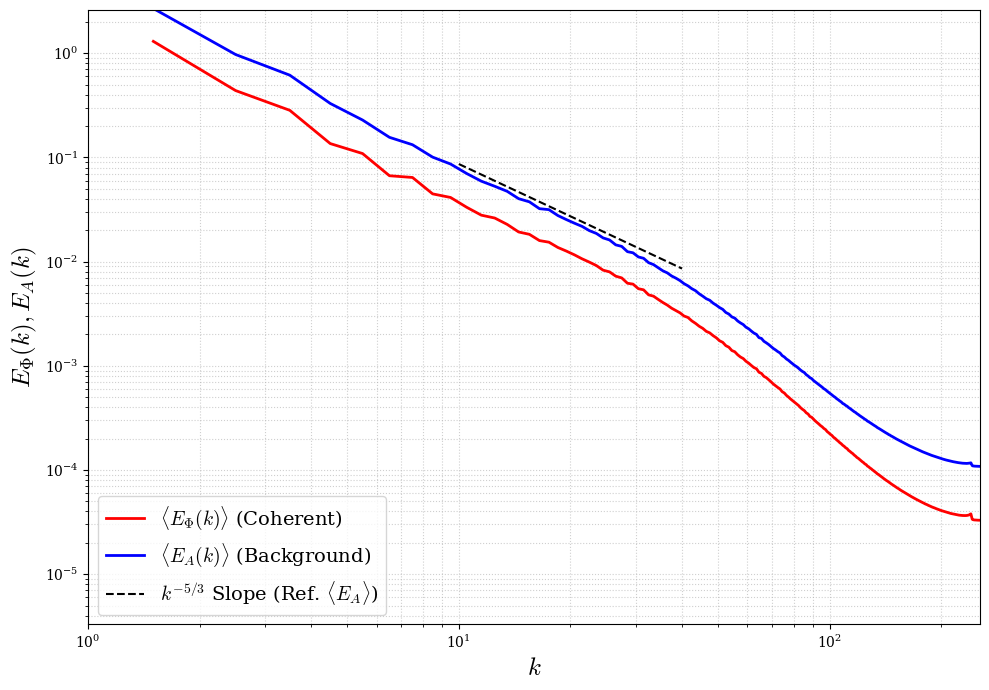}
    \caption{The ensemble-averaged decomposed edicity spectra for the coherent field, $\langle E_\Phi(k) \rangle$, and the background field, $\langle E_A(k) \rangle$. Both spectra exhibit the classical Kolmogorov $k^{-5/3}$ scaling, confirming that the coherent structures ($\Lvec_\Phi$) are self-similar and persist across the entire inertial cascade.}
    \label{fig:decomposed_spectra}
\end{figure}

Figure \ref{fig:decomposed_spectra} presents the ensemble-averaged edicity spectra, $\langle E_\Phi(k) \rangle$ and $\langle E_A(k) \rangle$ which shows that both the coherent and background spectra exhibit the classical Kolmogorov $k^{-5/3}$ power law with good agreement across the inertial range ($7 \lesssim k \lesssim 50$).\\

The fixed vertical separation between the two curves provides the empirical basis for testing the equipartition prediction. To quantify this partition precisely, we compute the Coherence Function, $\mathcal{R}(k)$, defined as the fractional contribution of the coherent field to the total edicity (see Eq. \ref{eq:R_k_def}).\\

\begin{figure}[h!]    
    \centering
    % Left Plot: Unwindowed
    \begin{subfigure}{0.48\textwidth}
        \centering
        \includegraphics[width=\linewidth]{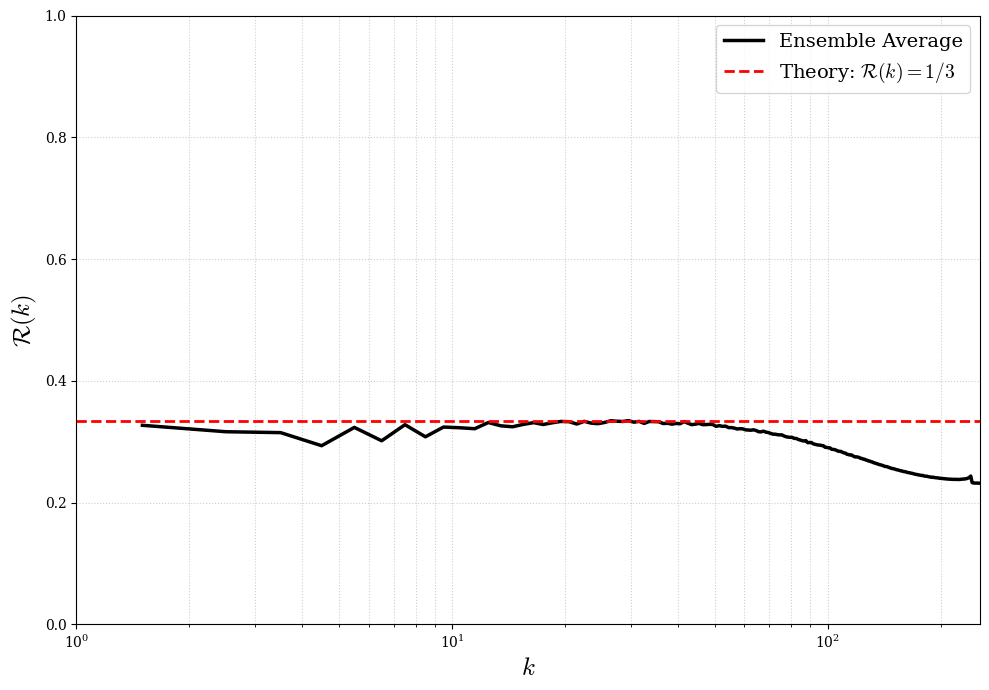}
        \caption{Unwindowed Evaluation}
        \label{fig:Rk_unwindowed}
    \end{subfigure}\hfill
    % Right Plot: Windowed
    \begin{subfigure}{0.48\textwidth}
        \centering
        \includegraphics[width=\linewidth]{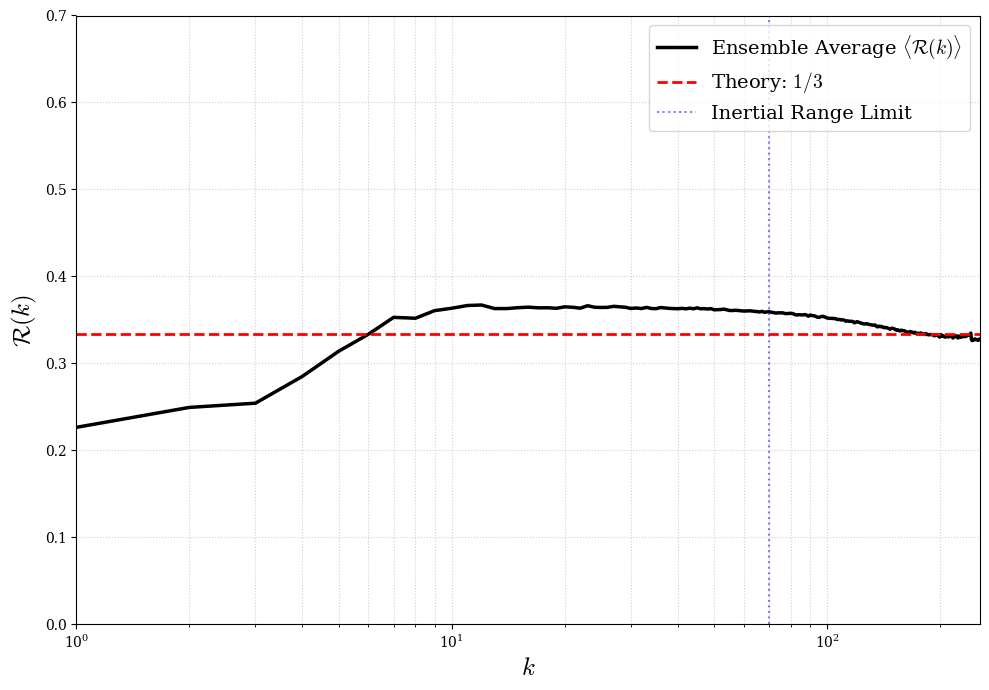}
        \caption{Windowed Evaluation}
        \label{fig:Rk_windowed}
    \end{subfigure}
    
    \caption{The coherence function, $\mathcal{R}(k)$. (a) The unwindowed spectrum exhibits a pristine equipartition plateau at the theoretical continuum limit of $1/3$. (b) The application of a spatial boundary window shifts the robust plateau to $\mathcal{R} \approx 0.36$ due to the artificial compressive source terms introduced by the finite control volume. Together, these evaluations validate the statistical prediction that the inertial range equipartitions energy according to the 1:2 ratio of the topological degrees of freedom.}
    \label{fig:coherence_function}
\end{figure}

Figure \ref{fig:coherence_function} displays a robust plateau at $\mathcal{R}(k) \approx 34-0.36$ throughout the inertial range confirming the statistical prediction $\langle E_A \rangle = 2 \langle E_\Phi \rangle$. The stability of this ratio indicates that the turbulent state operates at a specific statistical equilibrium where the energy of the background fluctuations is exactly double the energy of the structural order, governed strictly by the ergodic exploration of the phase space volume.\\

Figure \ref{fig:coherence_function}  presents the spectral edicity partition ratio, $R(k)$, evaluated both with and without the application of a spatial boundary window. While the unwindowed evaluation exhibits a pristine equipartition plateau at the theoretical continuum limit of $1/3$, applying a Super-Gaussian spatial window $W(r)$ is formally required to strictly enforce periodicity on a finite computational sub-domain. \\

However, we must explicitly acknowledge that this windowing operation inevitably introduces a deterministic spectral distortion. The topological separation relies on the divergence of the field, which for the windowed angular momentum transforms as:

\begin{equation}
\nabla \cdot (W \Lvec) = W(\nabla \cdot \Lvec) + \nabla W \cdot \Lvec
\end{equation}

Because the spatial gradient of the window ($\nabla W$) is strictly non-zero near the domain boundaries, the cross-term $\nabla W \cdot \Lvec$ acts as an artificial compressible source term. This mathematical artifact systematically injects artificial longitudinal energy into the coherent potential field ($\Phi_L$), inflating $E_\Phi$ and shifting the baseline of $R(k)$ upward (yielding the observed plateau at $R(k) \approx 0.36$). 

Therefore, the unwindowed spectrum reveals the true canonical equipartition of the unperturbed inertial cascade, while the windowed spectrum represents a necessary but distorting convolution of this physical equilibrium with the geometric gradient signature of the numerical control volume.

\subsection{Dynamics of the Edicity Cascade: The Entropic Flux}

Noting the static 1:2 partition of edicity, we verify the dynamical mechanism that sustains this equilibrium. The statistical theory predicts that the background field acts as a thermal bath for the coherent structures, implying a continuous, irreversible flux of edicity from $\Phi_L$ to $\AL$ . This flux is quantified by the transfer term $\Ttransfer$ (defined by Eq. \ref{eq:edicity_transfer}).\\

We calculated the instantaneous volume-integrated transfer rate, $\mathcal{T}_{\Phi \to A}$, across an ensemble of 20 distinct time snapshots to ensure statistical convergence. The results are summarized in Table \ref{tab:transfer_rate_ensemble}.\\

\begin{table}[h!]
\centering
\caption{Instantaneous and Ensemble-Averaged Transfer Rate ($\mathcal{T}_{\Phi \to A}$) Measured Across 20 Time Snapshots.}
\label{tab:transfer_rate_ensemble}
\begin{tabular}{cc}
\toprule
\textbf{Time, $t$} & \textbf{Instantaneous $\mathcal{T}_{\Phi \to A}$} \\
\midrule
101  & $3.3516  \times 10^{-3}$  \\
201  & $1.3759  \times 10^{-2}$  \\
301  & $1.3944  \times 10^{-2}$  \\
401  & $1.2738  \times 10^{-2}$  \\ 
501  & $4.1063  \times 10^{-3}$  \\
601  & $1.4471 \times 10^{-3}$  \\
701  & $8.1276  \times 10^{-3}$  \\
801  & $1.0179  \times 10^{-2}$  \\
901  & $1.8515  \times 10^{-2}$  \\
1001 & $1.9748  \times 10^{-2}$  \\
1101 & $1.9566  \times 10^{-2}$  \\
1201 & $3.0537  \times 10^{-2}$  \\
1301 & $3.7443  \times 10^{-2}$  \\
1401 & $3.5701  \times 10^{-2}$  \\
1501 & $3.0205  \times 10^{-2}$  \\ 
1601 & $3.0642  \times 10^{-2}$  \\
1701 & $3.3680  \times 10^{-2}$  \\
1801 & $2.1369  \times 10^{-2}$  \\
%2001 & $2.0973  \times 10^{-3}$  \\
\midrule
\textbf{Ensemble Average} & \textbf{$1.906 \times 10^{-2}$} \\
\bottomrule
\end{tabular}
\end{table}

The ensemble average is strictly positive ($\langle \Ttransfer \rangle > 0$). This confirms that the net flow of edicity is consistently from the coherent structures ($\Lvec_\Phi$) to the background field ($\Lvec_A$), validating the thermodynamic interpretation of the cascade as a relaxation process from low-entropy structure to high-entropy noise.\\

\begin{figure}[h!]
    \centering
    \begin{subfigure}[b]{0.48\textwidth}
        \includegraphics[width=\textwidth]{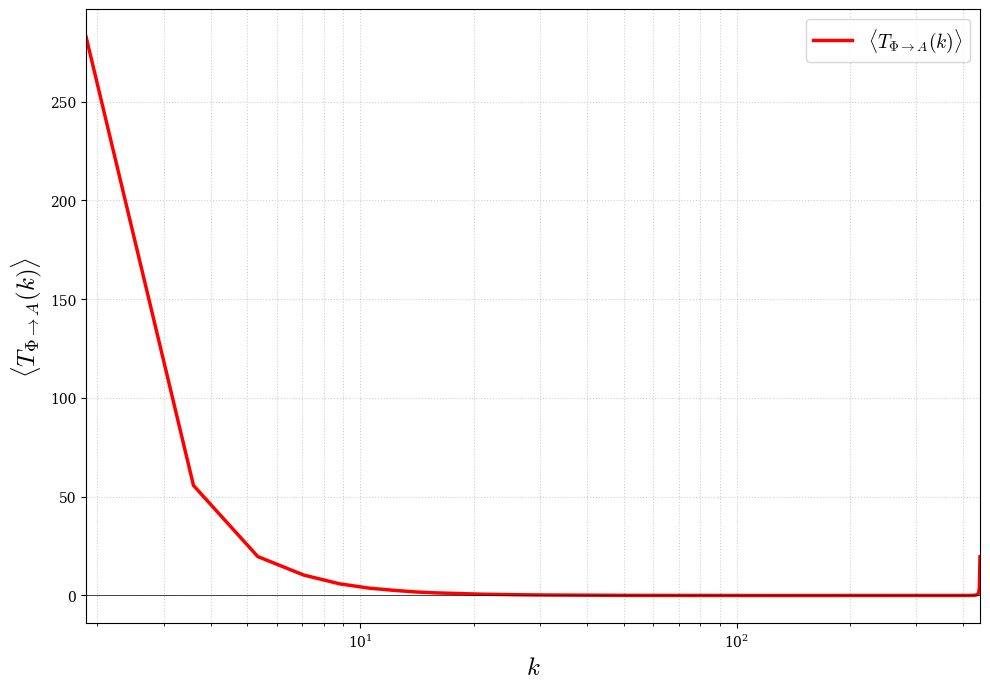}
        \caption{Spectral Transfer $\hatTtransfer(k)$}
    \end{subfigure}
    \hfill
    \begin{subfigure}[b]{0.48\textwidth}
        \includegraphics[width=\textwidth]{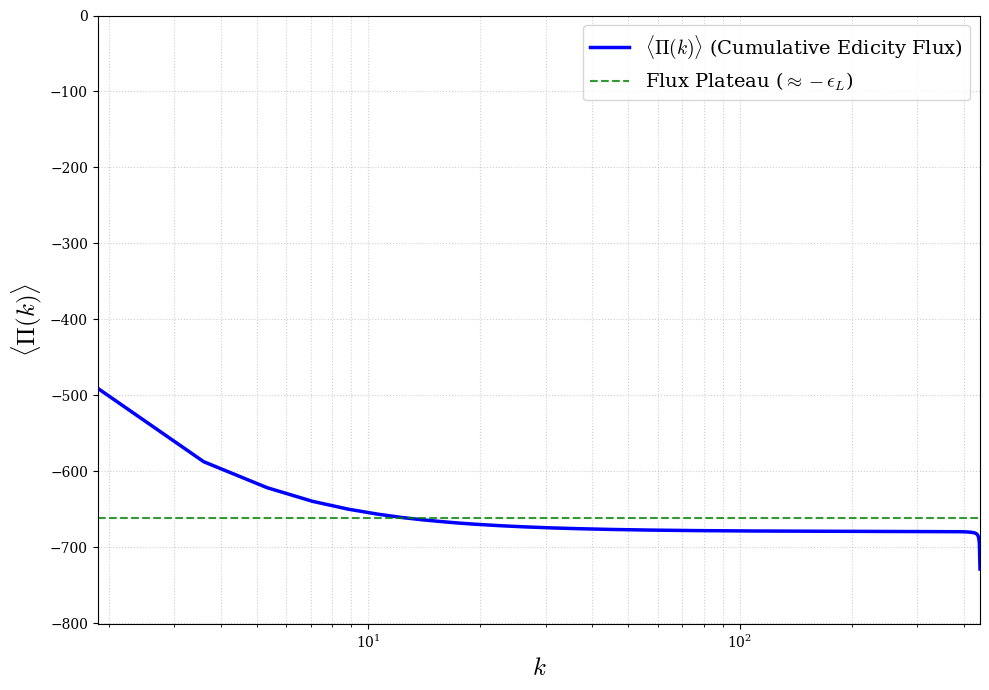}
        \caption{Cumulative Flux $\Pi(k)$}
    \end{subfigure}
    \caption{The dynamical driver of the edicity cascade. (a) The transfer spectrum is positive at low wavenumbers, indicating energy injection by large-scale coherent structures. (b) The cumulative flux exhibits a constant plateau in the inertial range, satisfying the stationarity condition for a cascade.}
    \label{fig:edicity_transfer}
\end{figure}

To resolve this dynamics across scales, we examine the spectral transfer function $\Ttransfer(k)$ (defined by Eq. \ref{eq:spectral_Ttransfer}) and its cumulative flux $\Pi(k)$ (defined by Eq. \ref{Pi_def}) , shown in Figure \ref{fig:edicity_transfer}.\\

The spectral transfer (Figure \ref{fig:edicity_transfer}a) is positive at large scales, confirming that coherent structures are the source of the cascade. The cumulative flux (Figure \ref{fig:edicity_transfer}b) exhibits a clean plateau throughout the inertial range. This constant flux confirms that the 1:2 partition is not a static pile-up of energy, but a dynamic steady state maintained by the scale-invariant hand-off of angular momentum.

\subsection{Verification of Geometric Quantization and the Partition of Unity}

We verify the physical quantization of the velocity field itself where the total kinetic energy is distributed among three distinct modes: the Radial strain mode ($\uvec_r$), the Coherent tangential mode ($\uvec_\Phi$), and the Background tangential mode ($\uvec_A$).\\

\begin{figure}[h!]
    \centering
    \begin{subfigure}[b]{0.48\textwidth}
        \includegraphics[width=\textwidth]{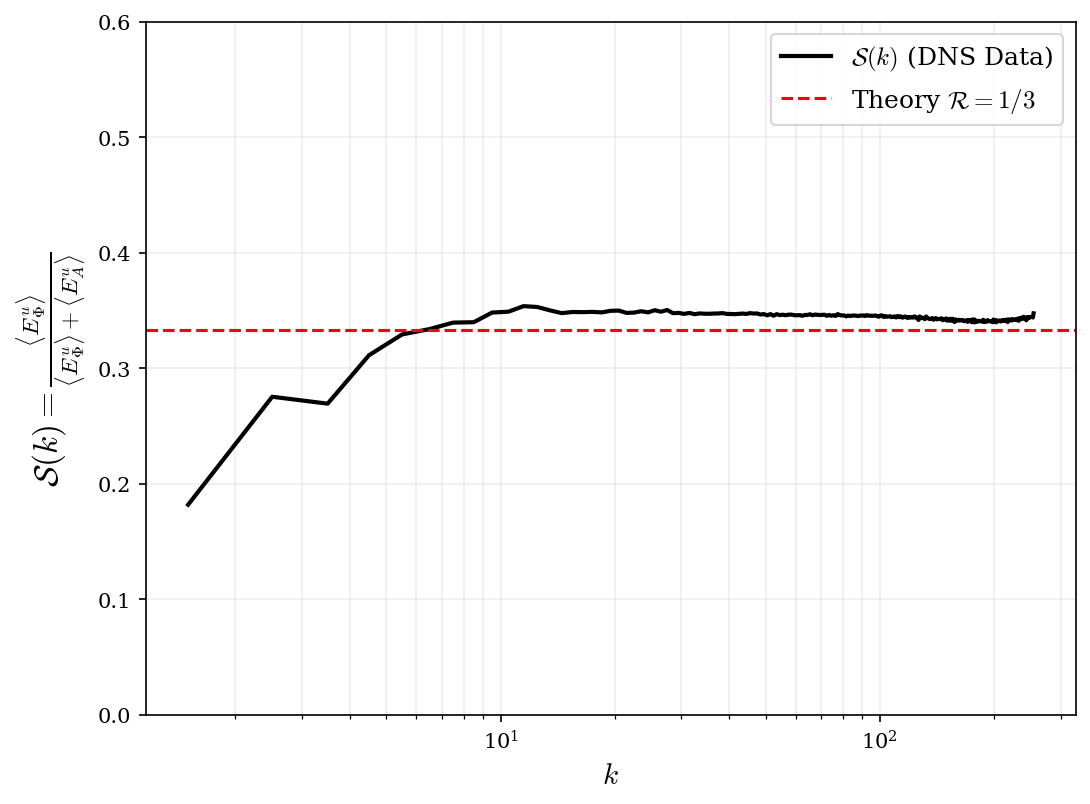}
        \caption{Tangential Partition $\mathcal{S}(k)$}
    \end{subfigure}
    \hfill
    \begin{subfigure}[b]{0.48\textwidth}
        \includegraphics[width=\textwidth]{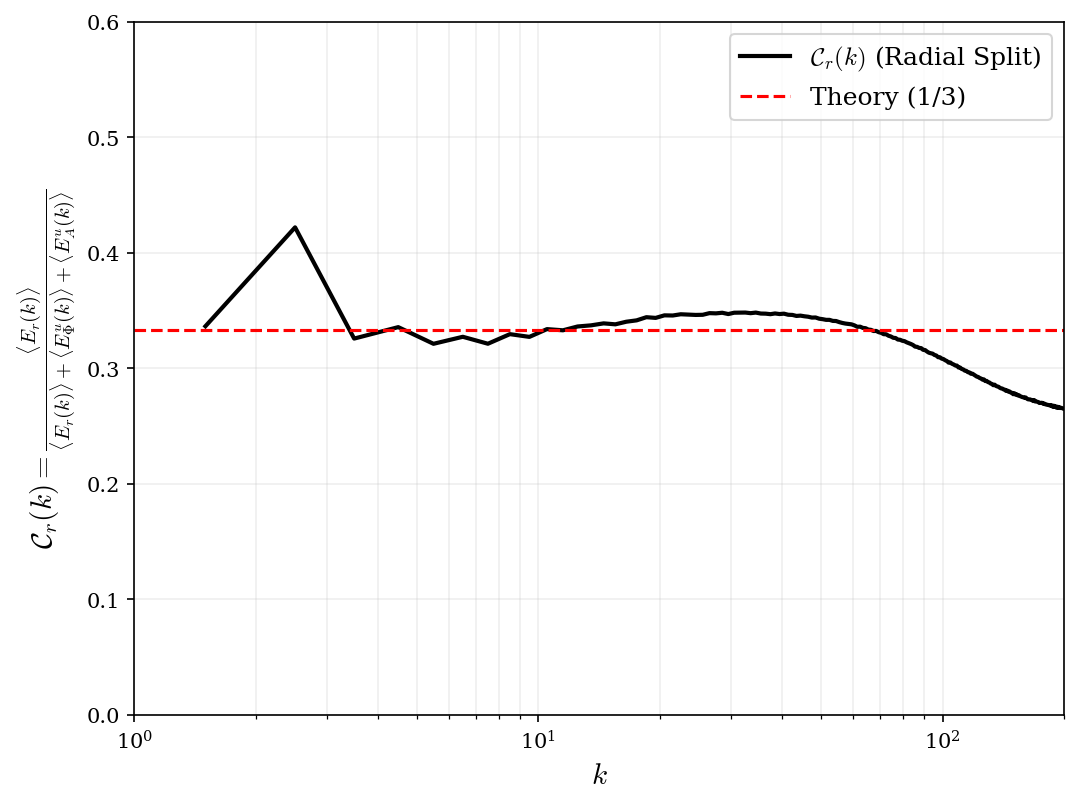}
        \caption{Radial Partition $\mathcal{C}_{r}(k)$}
    \end{subfigure}
    \caption{Verification of the Geometric Partition of Unity. (a) The tangential velocity field maintains a structural ratio of $\mathcal{S} \approx 0.35$, consistent with the 1:2 prediction for coherent vs. background modes. (b) The total velocity field maintains a radial energy fraction of $\mathcal{C}_{r} \approx 0.36$, consistent with the 1/3 geometric prediction for the radial strain mode. Together, these plateaus confirm that the turbulent cascade organizes itself according to the available degrees of freedom ($1 : 2 : 3$).}
    \label{fig:velocity_coherence}
\end{figure}

To validate this, we examine the scale-dependent coherence functions for both the tangential and radial partitions.\\

Figure \ref{fig:velocity_coherence} presents the two fundamental velocity ratios. Panel (a) displays the Velocity Coherence Function, $\mathcal{S}(k)$, which measures the structural fraction of the tangential energy. The data exhibits a robust plateau at $\mathcal{S} \approx 0.35$, confirming that the 1:2 partition between coherent structures and the background bath persists in velocity space.\\

Panel (b) displays the radial coherence function, $\mathcal{C}_{r}(k) = \langle E_r(k) \rangle / \langle E_{total}(k) \rangle$ (see Eq. \ref{eq:ratio_radial}). This metric quantifies the dimensionality constraint—the fraction of energy that must reside in the radial breathing mode to sustain a 3D flow. The data reveals a remarkably stable plateau at $\mathcal{C}_{r} \approx 0.35$ throughout the inertial range, closely approximating the geometric prediction of $1/3$.\\

The convergence of these three independent statistical measures—$\mathcal{R} \approx 1/3$ for edicity, $\mathcal{S} \approx 1/3$ for tangential velocity, and $\mathcal{C}_{r} \approx 1/3$ for radial energy—serves as a "triangulation" of the underlying physics. It confirms that the partition numbers are not artifacts of a specific coordinate system, but represent a fundamental geometric equilibrium of the Navier-Stokes equations.\\

\begin{figure}[h!]
    \centering
    % Adjusted width to 0.8\textwidth for a standalone figure. 
    % Change to 0.48\textwidth if you still want it small.
    \includegraphics[width=0.5\textwidth]{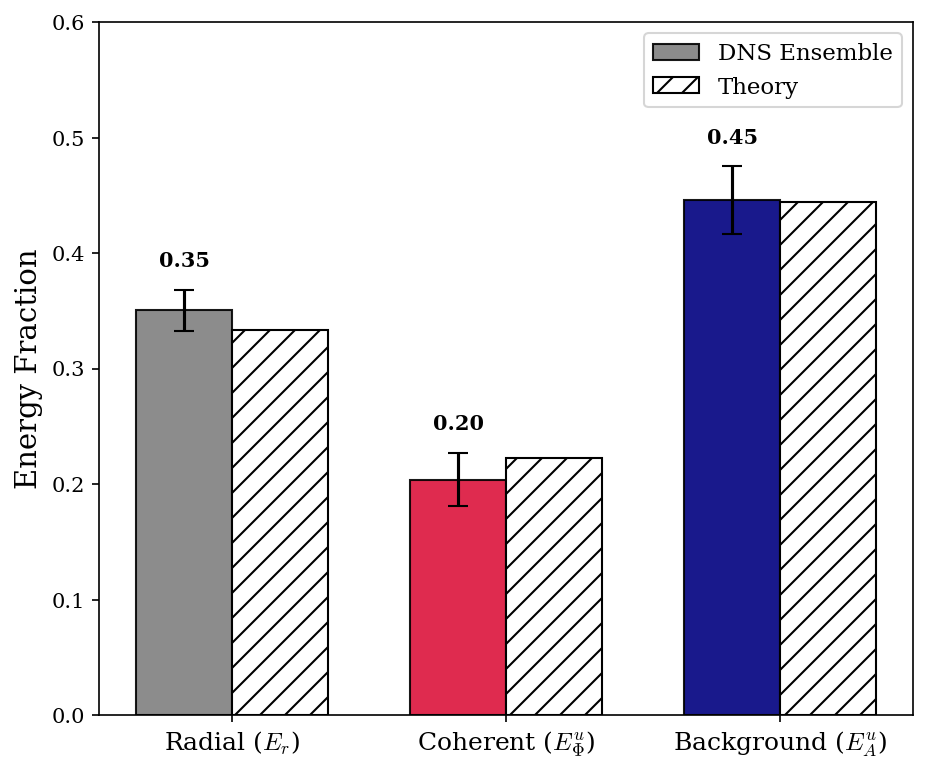}
    
    % Caption updated to contain only the text regarding the Grand Partition
    \caption{The partitioning of turbulent kinetic energy. The measurements confirm the predicted hierarchy, with the radial energy accounting for approximately 1/3 of the total flow.}
    
    \label{fig:grand_partition}
\end{figure}

Figure \ref{fig:grand_partition}  presents the ensemble-averaged global energy fractions in physical (not spectral) space (see Eqs. \ref{eq:ratio_radial}, \ref{eq:ratio_A}, \ref{eq:ratio_phi}). The data accurately confirms the predicted hierarchy $E_A > E_r > E_\Phi$. The background energy $E_A$ matches the theoretical prediction of $4/9 \approx 0.44$ almost exactly, confirming that the solenoidal background field captures the robust statistical backbone of the tangential flow. The Radial energy $E_r$ is measured at $\approx 0.35$, slightly exceeding the theoretical equipartition value of $1/3 \approx 0.33$. This surplus is enforced by the dimensionality constraint required to sustain the flow because the radial strain mode acts as the active driver of the cascade (quantified subsequently by the transfer spectrum in Fig.~\ref{fig:transfer_engine}), it maintains a slightly higher energy density than the passive tangential modes in a finite Reynolds number flow.\\

\begin{figure}[h!]
    \centering
    \includegraphics[width=0.7\textwidth]{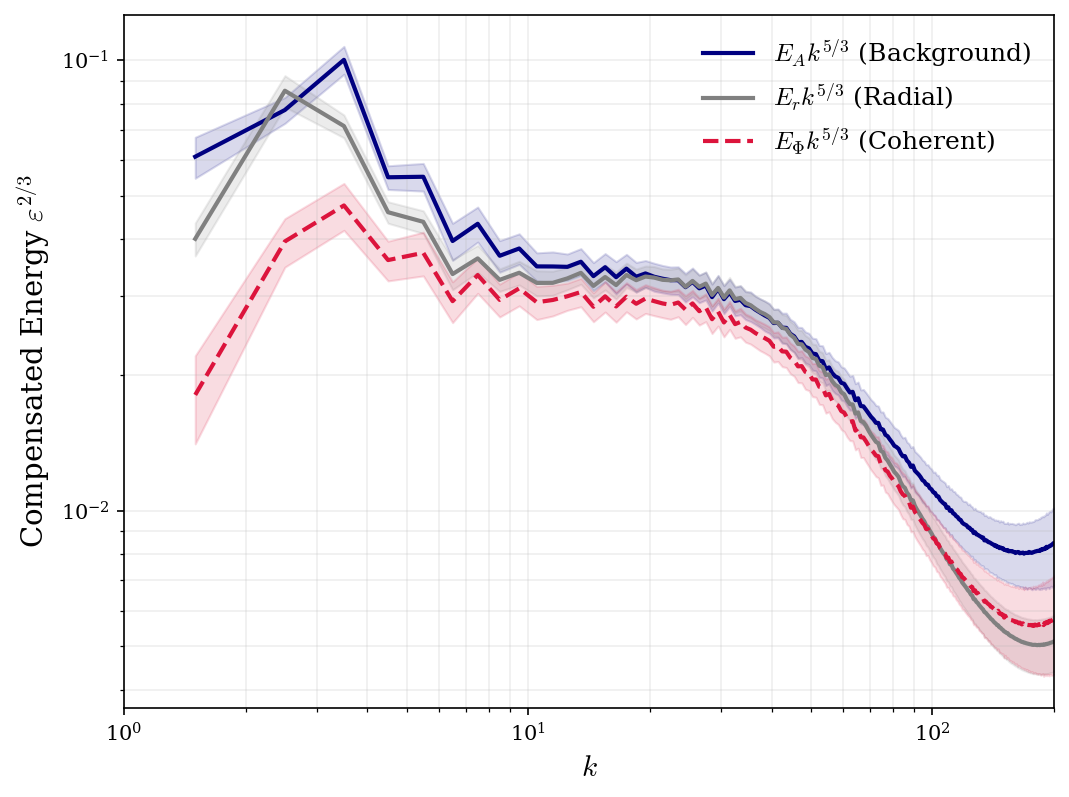}
    \caption{The Spectral Trinity (Compensated) of $E^u_\Phi k^{5/3}$, $E^u_A(k)k^{5/3}$ and $E_r^u(k)k^{5/3}$. The parallel plateaus confirm that all three components—Radial (gray), Coherent (red), and Background (blue)—independently obey Kolmogorov scaling. The crossover at $k \approx 12-13$ marks the transition from the energy-containing range (dominated by tangential sweeping) to the dissipation range (dominated by radial strain).}
    \label{fig:spectral_trinity}
\end{figure}

To confirm that these components are dynamic features of the cascade rather than static artifacts, we examine their spectral scaling. Figure \ref{fig:spectral_trinity} displays the compensated spectra for all three components: $E_r(k) k^{5/3}$, $E_\Phi(k) k^{5/3}$, and $E_A(k) k^{5/3}$.\\

The emergence of three distinct, parallel plateaus in the inertial range ($10 \lesssim k \lesssim 30$) is the definitive confirmation of the theory. It demonstrates that the  radial energy and the coherent energy are active dynamical participants in the Kolmogorov cascade. If the radial mode were merely a geometric artifact of the projection, it would exhibit noise-like scaling ($k^0$ or $k^{-1}$); instead, it obeys the $k^{-5/3}$ law of inertial turbulence.\\

Notably, we observe a crossover behavior at $k \approx 15$. At large scales ($k < 15$), the background rotation $E_A$ dominates whereas at small scales ($k > 15$), the radial strain energy $E_r$ rises relative to the background, crossing over to become the dominant sub-component. This shift is a sign of intermittency, where intense strain events dominate the dissipation range.\\

\subsection{The Energetic Cycle: Centrifugal Pumping and Vortex Stretching}

The static partition of energy describes the equilibrium state of the turbulence, but it does not explain how this state is sustained against dissipation. To isolate the dynamical mechanism driving the cascade, we compute the spectral transfer function $\langle \hat{T}_{r \to \tau}(k) \rangle$ (defined by Eq.~\ref{eq:T_tau_r}), which quantifies the rate of work done by the radial strain field on the tangential coherent structures.\\

\begin{figure}[h!]
    \centering
    \includegraphics[width=0.7\textwidth]{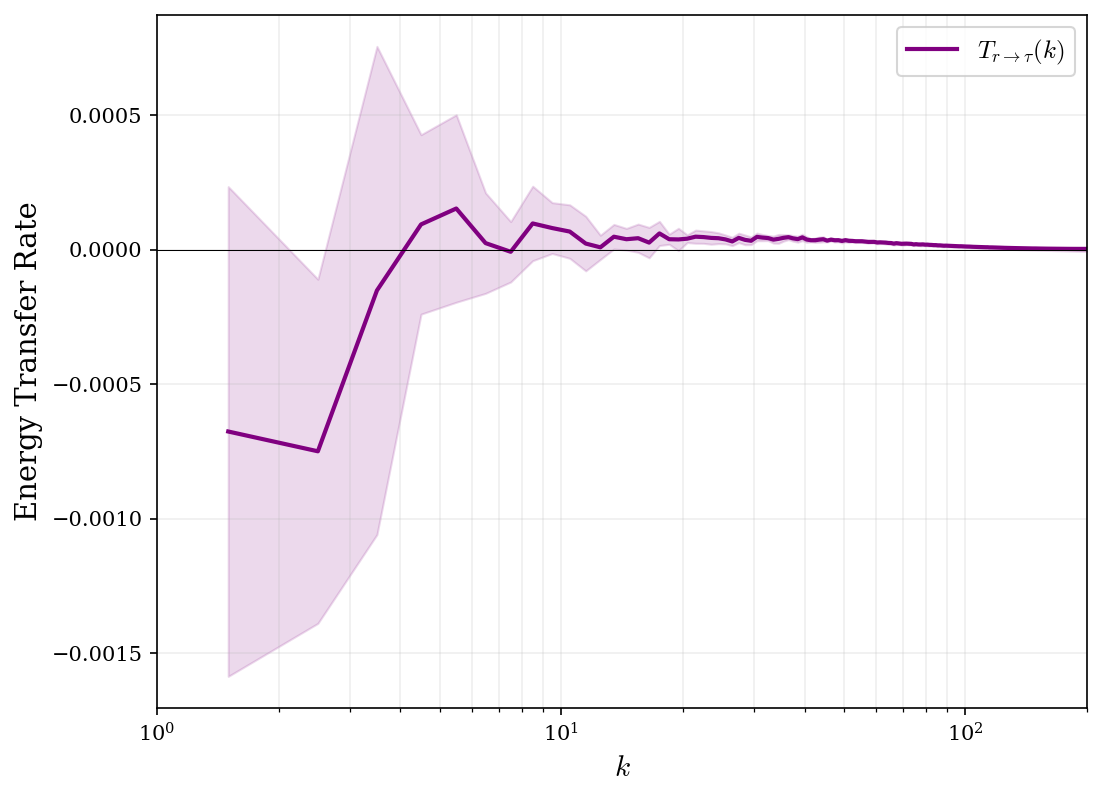}
    \caption{The driver of the Cascade. The spectral transfer function $T_{r \to \tau}(k)$ reveals the direction of the energy flux between the tangential and radial modes. The transition from negative (large scales) to positive (inertial range) confirms the cycle of centrifugal pumping followed by vortex stretching.}
    \label{fig:transfer_engine}
\end{figure}

Figure \ref{fig:transfer_engine} presents the ensemble-averaged transfer spectrum. The data reveals the dual-signed structure that maps the life cycle of turbulent energy:\\

For large scales, ($k < 5$), the transfer is negative ($T_{r \to \tau} < 0$). Physically, this indicates a flux of energy from the tangential field to the radial field. We identify this as Centrifugal Pumping where the rotation of large, energy-containing eddies generates centrifugal stress, which performs work to drive the radial expansion of fluid elements.\\
    
For larger wavenumbers in the inertial range, ($k > 10$), the transfer reverses sign and becomes positive ($T_{r \to \tau} > 0$). This indicates that the radial field is now performing work on the tangential field. We identify this as Vortex Stretching: the radial expansion field stretches the smaller tangential vortex tubes, intensifying their vorticity and injecting the energy required to sustain the forward cascade.\\

These measurements confirm the catalytic driver hypothesis derived from the tangentially projected Navier-Stokes equations (specifically the coupling term $\mathcal{T}_{\tau \to r}$ defined in Eq.~\ref{eq:T_tau_r}). Rather than acting as a passive reservoir; it acts as a catalytic intermediate, extracting energy from large-scale rotation via centrifugal force and reinjecting it as small-scale stretching work. This feedback loop is the dynamical driver that enforces the $k^{-5/3}$ scaling.

\subsection{Topological Structure and the Inversion of Enstrophy}

We examine the small-scale topology of the flow by analyzing the distribution of enstrophy (vorticity squared). While the energy spectrum is dominated by the volume-filling background modes, the enstrophy spectrum provides a sensitive probe of the intense, singular structures (vortex filaments) that characterize turbulent intermittency.\\

\begin{figure}[h!]
    \centering
    \includegraphics[width=0.7\textwidth]{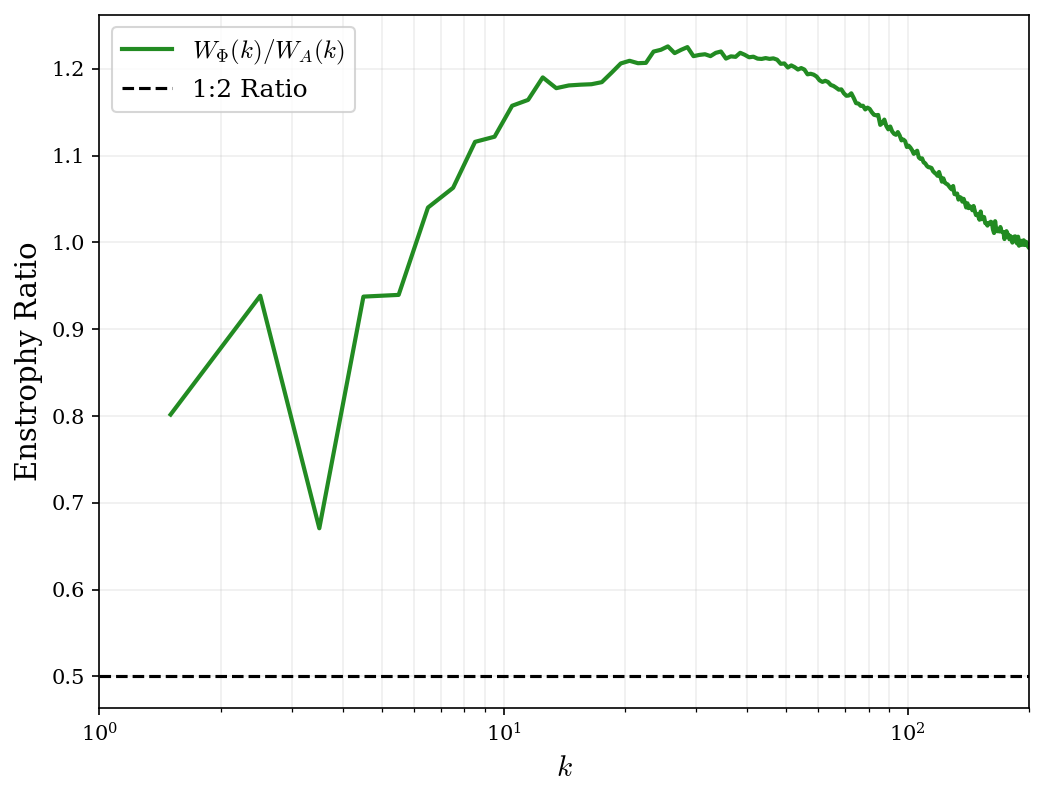}
    \caption{The Topological Inversion. The Enstrophy Ratio $W_\Phi(k)/W_A(k)$ reveals that while the background field contains most of the energy, the coherent field contains the majority of the small-scale enstrophy. The rise above unity at $k \approx 30$ confirms that $\uvec_\Phi$ captures the singular vortex filaments ("worms") responsible for dissipation.}
    \label{fig:enstrophy_inversion}
\end{figure}

Figure \ref{fig:enstrophy_inversion} presents the spectral Enstrophy Ratio, $W_\Phi(k) / W_A(k)$ where $W_\Phi(k)  = |\kvec \times \uvec_\Phi(k)|^2$ and $W_A(k)  = |\kvec \times \uvec_A(k)|^2$. A strict topological inversion is observed relative to the energy statistics. In the energy budget, the background field dominates the coherent field by a factor of two ($E_A \approx 2 E_\Phi$). In the enstrophy budget, however, this ratio rises sharply at high wavenumbers, reaching a peak of $\approx 1.2$ at $k \approx 30$.\\

This result indicates that the coherent potential $\Phi_L$ captures the intermittent, singular features of the flow. While the background field $\uvec_A$ represents the ``smooth,'' lower-gradient bath of turbulence, the coherent field $\uvec_\Phi$ resolves the intense, high-gradient vortex filaments (``worms'') that dominate the dissipation range.\\

Containing less than $25\%$ of the kinetic energy but over $50\%$ of the small-scale enstrophy, the coherent mode functions as the structural skeleton of the turbulence. The radial strain mechanism concentrates vorticity into these coherent filaments, establishing them as the primary loci for viscous dissipation (a dynamic captured analytically by the isolated viscous sink $\epsilon_\Phi$ derived in Eq.~\ref{eq:epsilon-Phi}). The kinematics are therefore fully coupled: the radial mode drives the cascade, the background mode stores the bulk thermalized energy, and the coherent mode provides the singular topology dictating ultimate dissipation.

\section{Mean Field Theory for Viscous Effects}

The Hamiltonian $H_0 = \frac{1}{2\rho} \int_V |\Lvec|^2 dV$, with the non-canonical Poisson bracket (Eq. \ref{eq:poisson_bracket}), exactly captures the nonlinear advective dynamics of the specific angular momentum $\Lvec$ in inviscid, incompressible isotropic turbulence, as has been shown previously. The energy transfer $\Ttransfer(k)$ between coherent ($\Lphi = -\nabla \PhiL$) and background ($\LA = \nabla \times \AL$) fields is intrinsic to the bracket, rendering an interaction Hamiltonian redundant. However, viscous effects in the $\Lvec$-transport Eq. \ref{eq:L-transport-equation} are dissipative and non-Hamiltonian, requiring alternative modeling. To incorporate this essential physical process into our statistical framework, we employ a Mean Field Theory (MFT) approach.\\

In homogeneous isotropic turbulence, the ensemble-averaged angular momentum vanishes, $\langle \Lvec \rangle = 0$ with the result that the field is entirely dominated by fluctuations, $\Lvec = \Lvec'$, and the mean viscous force is zero, $\langle \mu \nabla^2 \Lvec \rangle = 0$. However, the \textit{energetic} contribution of viscosity—the dissipation rate—is non-zero and positive.\\

The goal of the MFT approximation is to construct an \textit{effective} dissipative potential, $H_{\text{int}}$, that can be added to the Hamiltonian. This term is constructed such that, in the ensemble average, it correctly reproduces the energy damping effects of viscosity on the fluctuating modes. This effectively replaces the detailed, instantaneous viscous forces with a "mean field" damping parameter, analogous to the use of effective eddy viscosity in classical turbulence modeling.\\

\subsection{Modeling the Viscous Term via the Gradient Hamiltonian}

To incorporate viscosity, we introduce an effective dissipative Hamiltonian:

\begin{equation}
H_{\text{eff}} = H_0 + H_{\text{int}} 
\label{eq:H_visc_mf}
\end{equation}

where

\begin{equation}
H_{\text{int}} = \frac{\mu}{2\rho} \int_V |\nabla \Lvec|^2 dV
\end{equation}

$H_{\text{int}}$ approximates the dissipative effect of $\mu \nabla^2 \Lvec$. \\

To justify the inclusion of the gradient term in the effective Hamiltonian, we demonstrate that $H_{\text{int}} = \frac{1}{2} \int_V |\nabla \Lvec|^2 \, dV$ correctly models the viscous diffusion operator $\nabla^2 \Lvec$ in both a variational and energetic sense.

\subsubsection*{Variational Connection: The Diffusive Force}

The Hamiltonian $H_{\text{int}}$ acts as the potential energy functional for the field configuration. The "force" attempting to restore the field to a lower energy state is given by the negative functional derivative.

We consider a small perturbation $\delta \Lvec$ to the field. The variation in energy is:
\begin{equation}
\delta H_{\text{int}} = \frac{1}{2} \int_V \delta \left( \nabla \Lvec : \nabla \Lvec \right) \, dV = \int_V \nabla \Lvec : \nabla (\delta \Lvec) \, dV
\end{equation}
Using Green's First Identity (integration by parts) and assuming vanishing boundary terms (valid for periodic or unbounded domains):
\begin{equation}
\int_V \nabla \Lvec : \nabla (\delta \Lvec) \, dV = \oint_{\partial V} (\nabla \Lvec \cdot \mathbf{n}) \cdot \delta \Lvec \, dS - \int_V (\nabla^2 \Lvec) \cdot \delta \Lvec \, dV
\end{equation}
The surface integral vanishes, leaving:
\begin{equation}
\delta H_{\text{int}} = - \int_V (\nabla^2 \Lvec) \cdot \delta \Lvec \, dV
\end{equation}
By definition of the functional derivative, $\delta H = \int \frac{\delta H}{\delta \Lvec} \cdot \delta \Lvec \, dV$, we identify:
\begin{equation}
\frac{\delta H_{\text{int}}}{\delta \Lvec} = -\nabla^2 \Lvec
\end{equation}
Thus, the Laplacian $\nabla^2 \Lvec$ is the negative functional derivative of the gradient energy. Including this term in the Hamiltonian defines a system that evolves to minimize local gradients via diffusion.

\subsubsection*{Energetic Connection: The Dissipation Rate}
The term also correctly quantifies the instantaneous energy dissipation rate. The rate at which the viscous term $\mu \nabla^2 \Lvec$ removes energy from the system is:

\begin{equation}
D_L = \int_V \Lvec \cdot (\mu \nabla^2 \Lvec) \, dV
\end{equation}

Applying the vector identity $\nabla \cdot (\phi \mathbf{A}) = \nabla \phi \cdot \mathbf{A} + \phi \nabla \cdot \mathbf{A}$ to the components $L_i$:

\begin{equation}
\int_V L_i \nabla^2 L_i \, dV = \int_V \nabla \cdot (L_i \nabla L_i) \, dV - \int_V |\nabla L_i|^2 \, dV
\end{equation}

The divergence term vanishes by the Divergence Theorem (for periodic/unbounded domains). Summing over components gives:

\begin{equation}
D_L = -\mu \int_V |\nabla \Lvec|^2 \, dV = -2\mu H_{\text{int}}
\end{equation}

This confirms that $H_{\text{int}}$ is the precise measure of the capacity of the field configuration to dissipate energy via viscosity.

\subsection{Robustness of $\mathcal{R}(k)$}

To see that $\mathcal{R}(k)$ is robust even under the effects of viscosity, we consider $H_{\text{int}}$ in Fourier space. Since $\hat{\Lvec}(\kvec) = \hat{\Lphi}(\kvec) + \hat{\LA}(\kvec) = -i \kvec \hat{\Phi}_L(\kvec) + i \kvec \times \hatAL(\kvec)$, the gradient term maps to a wavenumber multiplication $\nabla \to i\kvec$. The interaction Hamiltonian now becomes:

\begin{equation}
H_{\text{int}} = \frac{\mu}{2\rho} \int \frac{d^3 k}{(2\pi)^3} k^2 \left( k^2 |\hat{\Phi}_L(\kvec)|^2 + |\kvec \times \hatAL(\kvec)|^2 \right).
\end{equation}

Combining this with the base Hamiltonian $H_0$, the effective Hamiltonian is:

\begin{equation}
H_{\text{eff}} = \int \frac{d^3 k}{(2\pi)^3} \left[ \frac{1}{2} \rho k^2 \left(1 + \frac{\mu}{\rho}k^2\right) |\hat{\Phi}_L(\kvec)|^2 + \frac{1}{2} \rho k^2 \left(1 + \frac{\mu}{\rho}k^2\right) |\kvec \times \hatAL(\kvec)|^2 \right].
\end{equation}

Here, the term $\frac{\mu}{\rho}k^2$ represents the ratio of the viscous relaxation rate to the inertial timescale at scale $k$. The partition function is:

\begin{equation}
\Zcal_{\text{eff}} = \int \mathcal{D}[\PhiL] \mathcal{D}[\AL] \exp(-\beta H_{\text{eff}})
\end{equation}    

Following the Gaussian integration methods used previously, the partition function factorizes as:

\begin{equation}
\Zcal_{\text{eff}} = (\Zcal_\Phi')^3
\end{equation}

The ensemble-averaged energies maintain the ratio $\langle \EA \rangle = 2 \langle \Ephi \rangle$, precisely because the viscous correction factor $\left(1 + \frac{\mu}{\rho}k^2\right)$ scales uniformly for both the scalar ($\PhiL$) and vector ($\AL$) degrees of freedom.\\

The Laplacian term introduces scale-dependent damping ($k^2$), effectively modeling the viscous cutoff that suppresses high-wavenumber fluctuations. However, because this damping acts isotropically on the angular momentum field, it preserves the 1:2 equipartition $\Rcal(k) \approx 1/3$ throughout the inertial range (where $\frac{\mu}{\rho}k^2 \ll 1$) and even into the dissipation range, assuming the effective viscosity remains isotropic.

\section{Thermodynamics of the Turbulent Ensemble}

The exact evaluation of the partition function (Eq.~\ref{eq:Z_factored}) allows us to move beyond simple energy averages and construct a complete thermodynamic description of the turbulent field. By treating the chaotic background fluctuations as a thermal bath, we can define macroscopic state functions—Entropy, Internal Energy, and Free Energy—that characterize the statistical equilibrium of the inertial range.\\

We define the effective temperature $T_\text{eff}$ of the turbulence, as the parameter characterizing the intensity of the background fluctuations, such that $\beta = 1/(k_B T_{\text{eff}})$. The fundamental thermodynamic potential is the Helmholtz Free Energy, $\mathcal{F}$, defined via the partition function:

\begin{equation}
\mathcal{F} = -k_B T_{\text{eff}} \ln \Zcal
\end{equation}

Using the factorization property derived in Eq.~\ref{eq:Z_factored} ($\Zcal = \Zcal_\Phi \cdot \Zcal_A$), the total free energy naturally decomposes into coherent and background components:

\begin{equation}
\mathcal{F}_{\text{total}} = \mathcal{F}_\Phi + \mathcal{F}_A = (-k_B T_{\text{eff}} \ln \Zcal_\Phi) + (-k_B T_{\text{eff}} \ln \Zcal_A)
\end{equation}

Since the background partition function is the square of the coherent one ($\Zcal_A = \Zcal_\Phi^2$, see Eq. ~\ref{eq:partition_balance}), we immediately obtain a thermodynamic equipartition law for the free energies:

\begin{equation}
\mathcal{F}_A = 2 \mathcal{F}_\Phi
\label{eq:free_energy}
\end{equation}

This indicates that the "energetic cost" of maintaining the disordered background field is exactly double the cost of maintaining the ordered coherent structures.\\

The Entropy, $S$, quantifies the volume of phase space accessible to the system (the degree of disorder). It is derived from the free energy via the relation $S = -(\partial \mathcal{F} / \partial T_{\text{eff}})$.\\

Applying this to our decomposed potentials yields a parallel 1:2 relation for the entropy:
\begin{equation}
S_A = 2 S_\Phi
\end{equation}

This result provides an insightful physical interpretation of the 1:2 edicity partition. In the inertial range, the turbulent state is not maximally disordered (which would imply $E_\Phi \to 0$). Instead, it settles into a specific equilibrium where the entropy of the background fluctuations ($S_A$) is exactly double the entropy of the coherent structures\footnote{While a rigorous thermodynamic mapping of the topological phases remains phenomenological, if one considers the limiting case where the macroscopic coherence is entirely embodied by the Helmholtz free energy ($\mathcal{F} \approx \langle E_\Phi \rangle$) and the chaotic background is entirely entropic ($T_{\text{eff}}S \approx \langle E_A \rangle$), the $1:2$ kinematic equipartition implies a provocative Virial-like balance for the inertial range: $T_{\text{eff}}S \approx 2\mathcal{F}$. This heuristic suggests the cascade preserves the $\mathcal{R}=1/3$ ratio to minimize the generalized free energy of the vacuum.} ($S_\Phi$).\\

The system seeks to minimize its generalized Helmholtz free energy, defined for each phase as:

\begin{align}
\mathcal{F}_\Phi &= \langle E_\Phi \rangle - T_{\text{eff}} S_\Phi \\
\mathcal{F}_A &= \langle E_A \rangle - T_{\text{eff}} S_A
\end{align}

We are now in a position to establish the condition for equilibrium between the two phases.

\subsection{Phase Equilibrium and the Two-Fluid Thermalized State}

The establishment of the free energy partition given by Eq. \ref{eq:free_energy}, ($\mathcal{F}_A = 2\mathcal{F}_\Phi$) allows us to formally define the criteria for stability in the turbulent cascade. \\

In classical thermodynamics, the stable coexistence of two interacting phases requires the equality of their chemical potentials, defined as the Gibbs (or Helmholtz) free energy per mole or particle. In our field-theoretic framework, the interacting "phases" are the coherent structural skeleton ($\PhiL$) and the thermalized background bath ($\AL$), while the "particles" correspond to the independent topological degrees of freedom available to each Fourier mode (intuitively, this can be seen from the path integrals being integrated over $\int \mathcal{D}[\Phi]$ and $\int \mathcal{D}[A_{L1}]\mathcal{D}[A_{L2}]$).\\

In the context of statistical field theory, the fundamental extensivity of the system is governed not by discrete particle numbers, but by the dimensionality of the phase space at each wavenumber $\kvec$; therefore, the classical mole number $N$ maps strictly to the number of independent, quadratic degrees of freedom.\\

The Helmholtz-Hodge decomposition dictates the geometry of this phase space: the scalar potential $\PhiL$ is constrained to a single longitudinal dimension, possessing one degree of freedom ($N_\Phi = 1$). The Coulomb-gauged vector potential $\AL$ is constrained to the transverse plane, possessing two independent polarization degrees of freedom ($N_A = 2$). \\

We can therefore define the chemical potential ($\mu_i =\frac{\partial \mathcal{F}_i}{\partial N_i}$) of each turbulent fluid as its Helmholtz free energy per degree of freedom. Utilizing the free energy partition derived above, the chemical potentials of the two phases evaluate to:

\begin{align}
\mu_\Phi &= \frac{\mathcal{F}_\Phi}{N_\Phi} = \frac{\mathcal{F}_\Phi}{1} = \mathcal{F}_\Phi \\
\mu_A &= \frac{\mathcal{F}_A}{N_A} = \frac{2\mathcal{F}_\Phi}{2} = \mathcal{F}_\Phi
\end{align}

This yields the exact equality:

\begin{equation}
\mu_\Phi = \mu_A
\end{equation}

This result constitutes the strict thermodynamic condition for phase equilibrium. It demonstrates that the $1:2$ energetic partition is not merely a static geometric coincidence, but the signature of a dynamic, thermalized two-fluid state. If an external forcing were to inject excess enstrophy into the coherent field, raising its free energy such that $\mu_\Phi > \mu_A$, the system would experience a restoring "thermodynamic pressure." To minimize the free energy, the flow would undergo a continuous phase transition: the excess coherent structures would become unstable, breaking down via the radial strain engine and scattering their edicity into the background bath ($\Ttransfer > 0$) until the phase equilibrium $\mu_\Phi = \mu_A$ is restored. \\

\section{Conclusion}

This study establishes a topological duality within the structure of fully developed isotropic turbulence with a first-principles statistical field theory. By decomposing the angular momentum field ($\Lvec$) via exact Helmholtz projection operators, we have demonstrated that the fluid continuum spontaneously segregates into two orthogonally distinct thermodynamic phases: a longitudinal condensate of macroscopic coherent structures ($\PhiL$) and a transverse, volume-filling thermal bath ($\AL$). \\

By defining an idealized Hamiltonian for the inertial range, a partition function could be constructed revealing that the fields are mathematically bound to a strict $1:2$ equipartition of available phase space. By mapping the result back to the physical velocity field ($\uvec = \uvec_A + \uvec_\Phi + \uvec_r$), revealed the radial $\uvec_r$ velocity field which was hidden in the $\Lvec$ framework allowed us to peer deeper and show a recursive partitioning scheme for the kinetic energies in the ratio $E^u_r:E^u_\Phi:E^u_A = 1/3 : 2/9 : 4/9$. \\

This framework was rigorously validated against high-resolution DNS data. The emergence of parallel $k^{-5/3}$ plateaus for all three components and the confirmation of the precise energy partitions ($E_\Phi:E_A = 1:2$) and  ($E^u_r:E^u_\Phi:E^u_A =1/3 : 2/9 : 4/9$) provide strong empirical evidence that isotropic turbulence is not merely a chaotic state, but a state of Geometric Equilibrium. The flow does not simply mix randomly; it dynamically adjusts its structure until it satisfies the rigid equipartition constraints imposed by the operator ranks we have identified.\\

The thermodynamics implied by the partition function showed that the topological quantization mandates that the turbulent steady state is not merely a cascade of energy, but a true thermodynamic equilibrium governed by the equalization of chemical potential parity $\mu_\Phi = \mu_A$.\\

While this spectral equipartition successfully captures the global thermodynamic equilibrium of the continuous fields, the turbulent cascade remains an open, mechanically driven system. The continuous influx of kinetic energy via the radial strain field---and its subsequent localized dissipation---suggests that the physical-space interactions between the coherent structures and the transverse bath are governed by transient, highly non-equilibrium energy exchanges.\\

\bibliographystyle{siam}
\bibliography{references}

\end{document}